\documentclass[nofootinbib,preprintnumbers,superscriptaddress,notitlepage]{revtex4}

\usepackage{epsfig}

\begin{document}
\title{Scaling of the CKM Matrix in the 5D MSSM}

\author{A.~S.~Cornell}
\email[Email: ]{alan.cornell@wits.ac.za}
\affiliation{National Institute for Theoretical Physics; School of Physics, University of the Witwatersrand,
Wits 2050, South Africa}
\author{Aldo Deandrea}
\email[Email: ]{deandrea@ipnl.in2p3.fr}
\affiliation{Universit\'e de Lyon, F-69622 Lyon, France; Universit\'e Lyon 1, CNRS/IN2P3, UMR5822 IPNL, F-69622 Villeurbanne Cedex, France}
\author{Lu-Xin~Liu}
\email[Email: ]{luxin.liu9@gmail.com}
\affiliation{National Institute for Theoretical Physics; School of Physics, University of the Witwatersrand,
Wits 2050, South Africa}
\author{Ahmad Tarhini}
\email[Email: ]{tarhini@ipnl.in2p3.fr}
\affiliation{Universit\'e de Lyon, F-69622 Lyon, France; Universit\'e Lyon 1, CNRS/IN2P3, UMR5822 IPNL, F-69622 Villeurbanne Cedex, France}

\begin{abstract}
We discuss a five-dimensional Minimal Supersymmetric Standard Model compactified on a $S^1/Z_2$ orbifold, looking at, in particular, the one-loop evolution equations of the Yukawa couplings for the quark sector and various flavor observables. Different possibilities for the matter fields are discussed, that is, where they are in the bulk or localised to the brane. The two possibilities give rise to quite different behaviours. By studying the implications of the evolution with the renormalisation group of the Yukawa couplings and of the flavor observables we find that, for a theory that is valid up to the unification scale, the case where fields are localised to the brane, with a large $\tan\beta$, would be more easily distinguishable from other scenarios.
\end{abstract}

\date{17 January, 2012}
\preprint{LYCEN 2011-11, WITS-CTP-81}
\maketitle


\section{Introduction}\label{sec:1}

\par The masses of the quarks and charged leptons are determined in the Standard Model (SM) via Yukawa couplings to the Higgs boson. The origin of their structure (masses and mixing angles) has no explanation within the SM and presents as one of the major challenges for physics beyond the SM. Among these models those with extra spatial dimensions offer many possibilities for model building and TeV-scale physics scenarios which can now be explored or constrained at the Large Hadron Collider (LHC). Extra-dimensional models allow, for example, a way to generate electroweak symmetry breaking or supersymmetry breaking through the choice of appropriate boundary conditions (for a review see Ref.\cite{XDrev}). In addition, for the case of flat extra-dimensions, the presence of towers of excited Kaluza-Klein (KK) states induces a power-law enhancement of the gauge couplings, leading to possible low-scale unification \cite{DDG,TeVstrings}. This effect can be applied to other couplings, such as Yukawa couplings, giving an original way to generate mass hierarchies \cite{DDG,YukawaXD}. The study of the renormalisation group equations (RGEs) provides a way by which partial explorations of the physics implications at a high energy scale is possible, as the theories at asymptotic energies may reveal new symmetries or other interesting properties which may lead to deeper insights into the physical content. As such, in order to understand these as yet unexplained hierarchies of mixing angles and fermion masses, a great deal of work has gone into analysing the RGEs \cite{Cornell:2010sz,Cheng:1973nv, Babu:1987im, Sasaki:1986jv, Machacek:1983fi, Liu:2009vh, Balzereit:1998id, Kuo:2005jt}. Therefore the behaviour of the well known quark sector's flavor mixings in the charged current, as described by the Cabibbo-Kobayashi-Maskawa (CKM) matrix, shall be a focus of this paper.

\par By using the RGEs, we will study the asymptotic properties of the Lagrangian parameters like the Yukawa coupling constants and mixing angles, and explore the possibility of a model in which the CKM matrix might have a simple, special form at asymptotic energies. This quest will, however, be limited by the fact that extra-dimensional theories are only effective ones, limited by a cut-off in their physical description of fundamental phenomena. Therefore the following study can only be used as an indication of the behaviour of couplings and mixing parameters at an intermediate scale between the electroweak scale (at which these parameters are measured) and the higher scale at which the effective theory ceases to be valid. However, in the range of the LHC energies (of the order of a few TeV), and beyond, one can indeed test if the departure from the usual behaviour of the coupling evolution can be seen in precision flavor measurements. In order to discuss the implications of an effective five dimensional supersymmetric theory, we first recall the usual formalism in the SM, where the CKM matrix has four observable parameters including three mixing angles and one phase. In the standard parametrisation it has the form:
\begin{equation}
V_{CKM} = \left( {\begin{array}{ccc}
{{V_{ud}}}&{{V_{us}}}&{{V_{ub}}}\\
{{V_{cd}}}&{{V_{cs}}}&{{V_{cb}}}\\
{{V_{td}}}&{{V_{ts}}}&{{V_{tb}}}
\end{array}} \right) = \left( {\begin{array}{ccc}
{{c_{12}}{c_{13}}}&{{s_{12}}{c_{13}}}&{{s_{13}}{e^{ - i{\delta}}}}\\
{ - {s_{12}}{c_{23}} - {c_{12}}{s_{23}}{s_{13}}{e^{i{\delta}}}}&{{c_{12}}{c_{23}} - {s_{12}}{s_{23}}{s_{13}}{e^{i{\delta}}}}&{{s_{23}}{c_{13}}}\\
{{s_{12}}{s_{23}} - {c_{12}}{c_{23}}{s_{13}}{e^{i{\delta}}}}&{ - {c_{12}}{s_{23}} - {s_{12}}{c_{23}}{s_{13}}{e^{i{\delta}}}}&{{c_{23}}{c_{13}}}
\end{array}} \right) \; , \label{eqn:CKM}
\end{equation}
where $s_{12} = \sin\theta_{12}$, $c_{12} = \cos\theta_{12}$ etc. are the sines and cosines of the three mixing angles $\theta_{12}$, $\theta_{23}$ and $\theta_{13}$, and $\delta$ is the CP violating phase.

\par The aim of this paper is to therefore study these effects explicitly in the case of one extra-dimension within a minimal supersymmetric model, which will be referred to as 5D MSSM. Note that as pointed out in Ref.\cite{Bhattacharyya:2010rm} a supersymmetric and extra-dimensional extension to the SM can be considered simultaneously, with the advantages that though higher dimensional field theories are non-renormalizable (requiring the existence of a UV completion), superstring theories (which contain supersymmetry) provide the main hope in this context. In fact, such a point of view has already been explored through AdS/CFT correspondences in accounting for both ``little" and ``big" hierarchies \cite{Sundrum:2009gv}. Also embedding the SM in an extra-dimensional space-time does not stabilise instabilities due to quantum corrections in the scalar potential, whilst supersymmetry may ameliorate this. Finally, most 4-dimensional supersymmetric models lack a simple mechanism for supersymmetry breaking, something the extra-dimensions may offer. To be precise, we will discuss a five dimensional $\mathcal{N}=1$ supersymmetric model compactified on the $S^1/Z_2$ orbifold as a simple testing ground for the effects of the extra-dimension on the quark Yukawa couplings and the CKM matrix observables. We will not discuss in the present paper the RGE evolution of the sparticle mixing parameters sector. This interesting question will be explored elsewhere as it requires more detailed assumptions on the sparticle spectrum and corresponding soft supersymmetry breaking (SSB) parameters, where at least some SSB parameter matrices can be non-diagonal in the basis in which quarks are diagonal (this implies flavor-violating decays, giving rise to the well known flavor problem in the MSSM and its extensions; even without considering the RGE evolution of the SSB parameters).

\par The paper is organised as follows: In section \ref{sec:2} we shall develop our model and calculate the various beta functions required to determine the evolution equations for the Yukawa couplings. We shall look at two cases, where the matter fields are localised on the brane, and where they are free to propagate in the bulk. In section \ref{sec:3} we shall then determine the evolutions of fermion Yukawa couplings, and the CKM matrix which follows from them. In section \ref{sec:4} we then numerically analyse the evolution behaviours of the physical observables, including the Yukawa couplings, quark flavor mixings, and the CP violation Jarlskog parameter. Our results and conclusions are presented in section \ref{sec:5}.


\section{The 5D MSSM and beta functions}\label{sec:2}

\par In this Universal Extra-dimensional (UED) formalism the compactification $S_1/{Z_2}$ implies that the 5D Lorentz invariance is broken to the usual 4D one, however, a remnant of momentum conservation along the fifth coordinate implies that KK number is conserved at tree level and that KK parity is conserved at loop level. The KK-parity invariance has two main implications in phenomenology: the contributions of the KK modes to electroweak precision observables arise only through loops, and the exact KK-parity implies that the lightest KK mode is stable and can be a cold dark matter candidate.

\par To begin our calculation we note that the beta functions can be derived more easily in the superfield formalism, where we shall now briefly discuss $\mathcal{N}=1$ supersymmetry in a five-dimensional Minkowski space and its description in terms of 4D superfields. Note that more details of this approach can be found in Refs.\cite{Deandrea:2006mh, Bouchart:2011va, Flacke, Hebecker, XDSUSY}, where, as a matter of notation, we shall label our space-time coordinates by $(x^{\mu}, y)$.

\par
\begin{figure}[t]
\begin{center}
\includegraphics[width=0.25\textwidth]{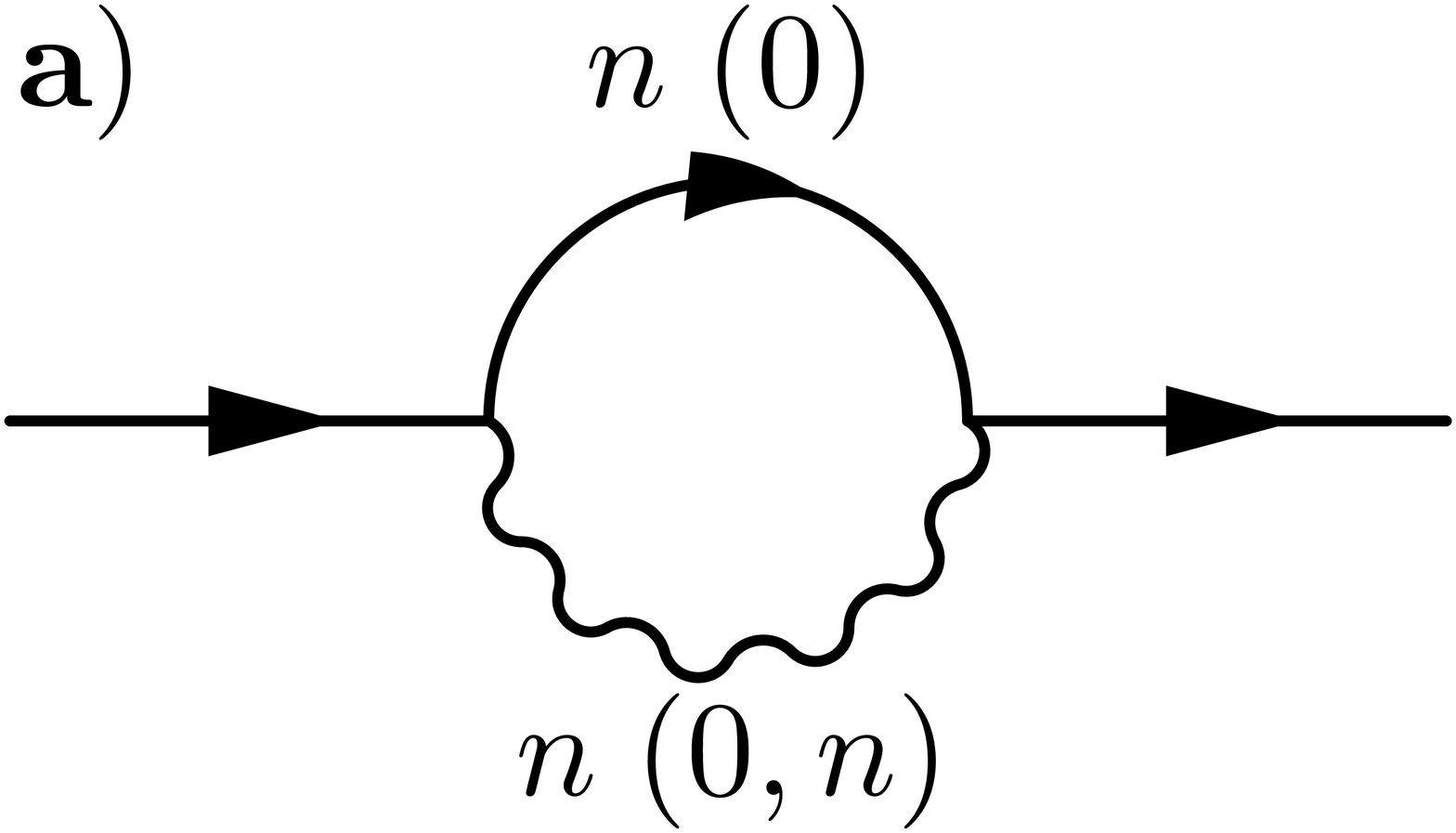}\hspace{0.05\textwidth}
\includegraphics[width= 0.25\textwidth]{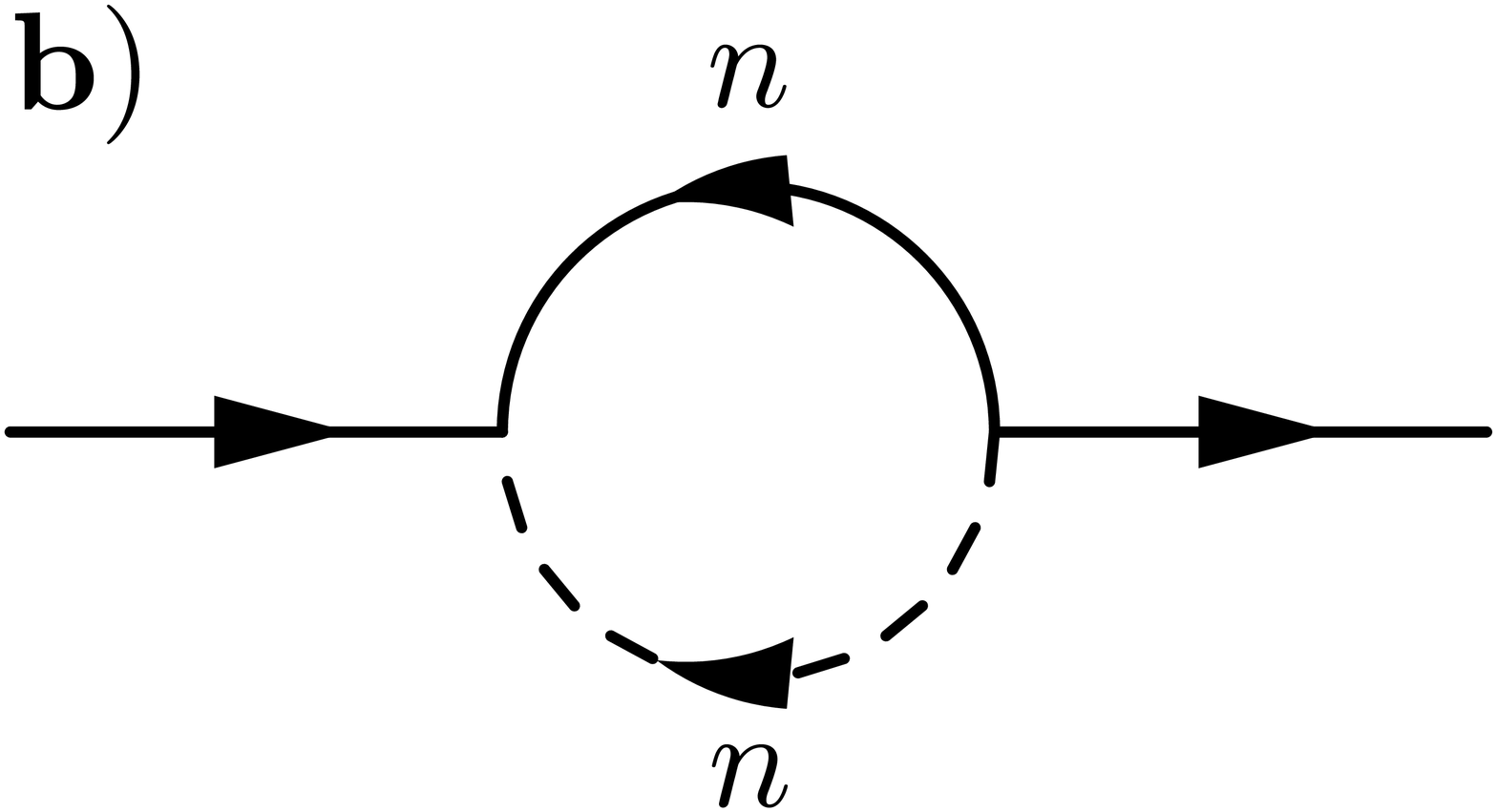}\hspace{0.05\textwidth}
\includegraphics[width= 0.25\textwidth]{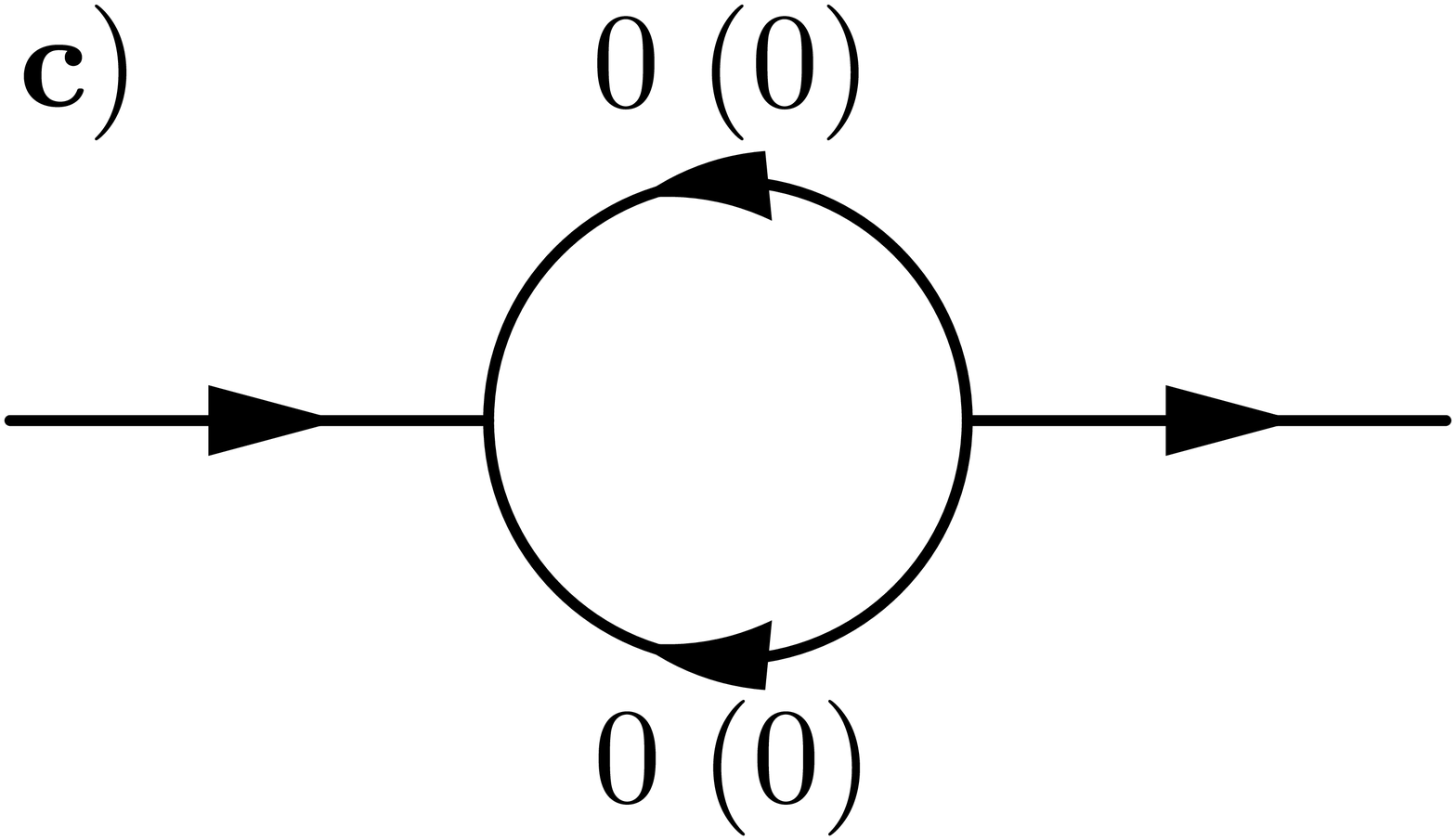}
\includegraphics[width= 0.25\textwidth]{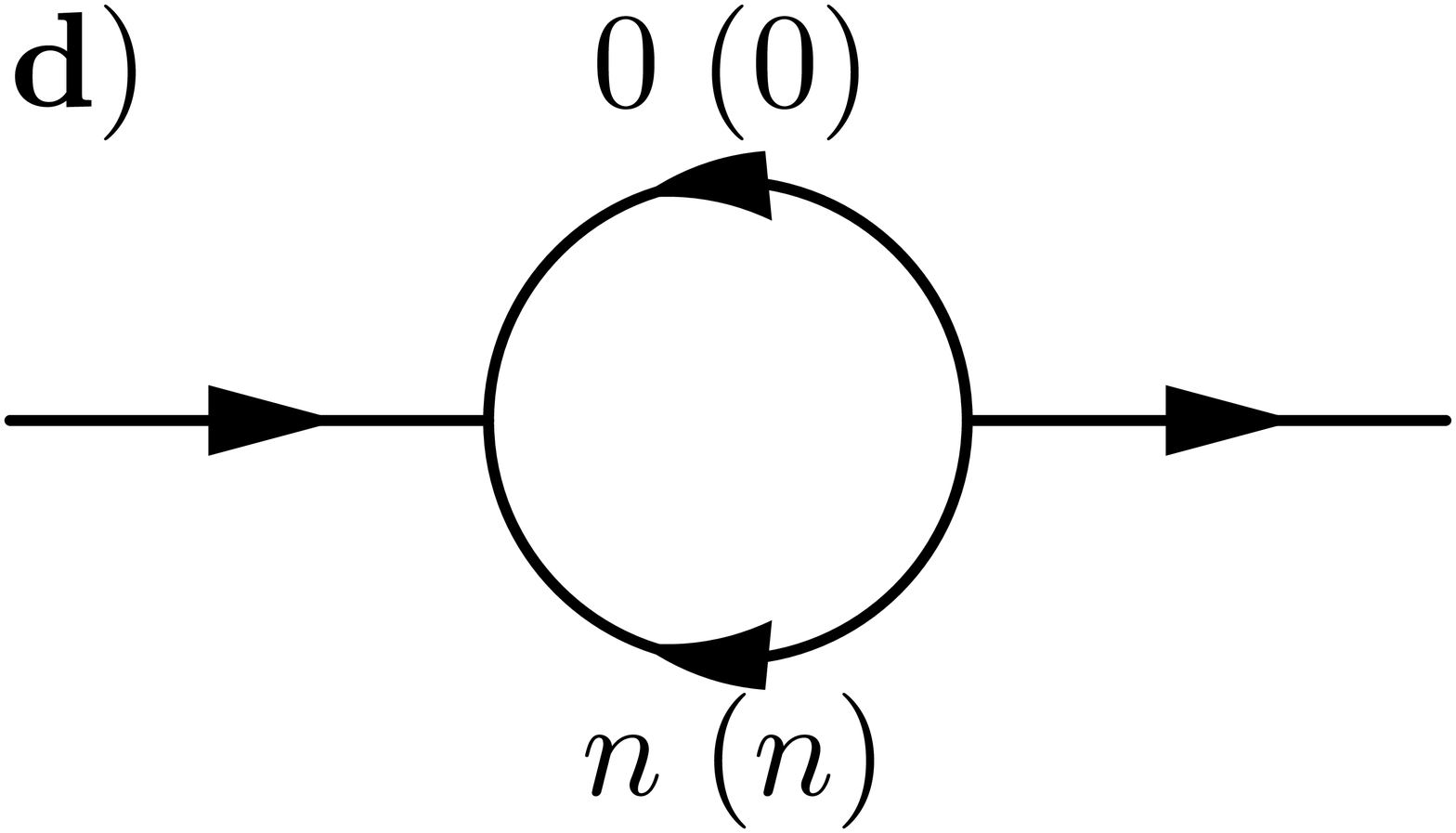}\hspace{0.05\textwidth}
\includegraphics[width= 0.25\textwidth]{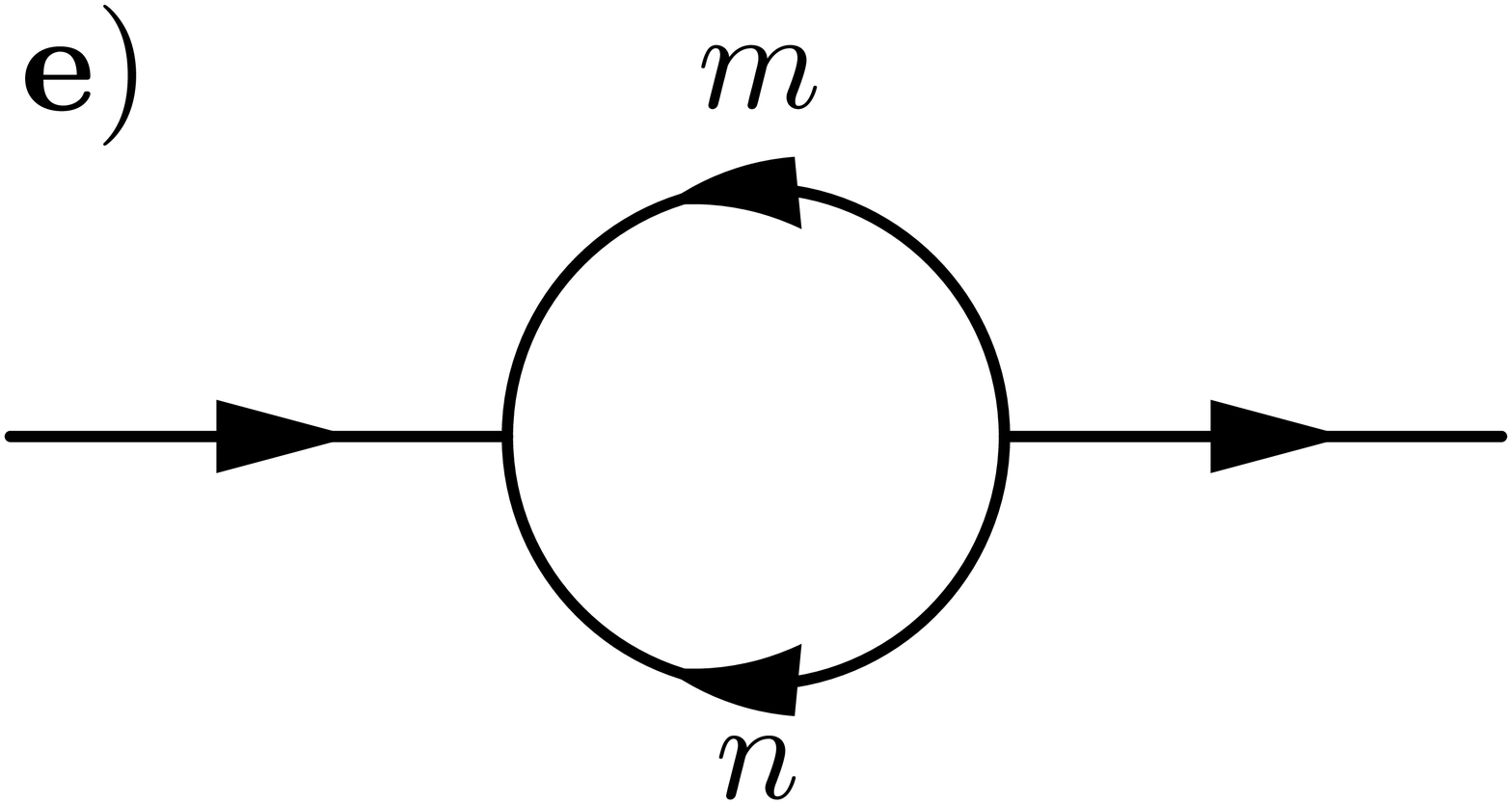}
\caption{\sl The one-loop diagrams related to the wave-function renormalisation of the matter superfields, in which diagrams a)-e) refer to the case where all the matter fields are in the bulk, and the excited KK states are labeled by the number without the bracket; whereas diagrams a), c) and d) are related to the brane localised matter fields case, in which the KK states are labeled by the number inside the bracket \cite{Deandrea:2006mh}.}\label{fig:1}
\end{center}
\end{figure}

 As an introductory example to this formalism, consider the gauge sector, described by a 5D $\mathcal{N}=1$ vector supermultiplet which consists (on-shell) of a 5D vector field $A^M$, a real scalar $S$ and two gauginos, $\lambda$ and $\lambda'$. The action for which can be given by:
\begin{eqnarray}
S_g &=& \int \mathrm{d}^5x\frac{1}{2kg^2}\mathrm{Tr}\left[-\frac{1}{2}F^{MN}F_{MN}-D^MSD_MS-i\overline{\lambda}\Gamma^MD_M \lambda -i\overline{\lambda}'\Gamma^MD_M\lambda'+(\overline{\lambda}+\overline{\lambda}')[S,\lambda+\lambda']\right] \; , \label{eqn:0}
\end{eqnarray}
with $D_M=\partial_M+iA_M$ and $\Gamma^M=(\gamma^\mu,i\gamma^5)$. $F^{MN}=-\frac{i}{g}[D^M,D^N]$ and $k$ normalises the trace over the generators of the gauge groups. One can also write these fields in terms of a $\mathcal{N}=2$, 4D vector supermultiplet, $\Omega=(V\,,\chi)$, where $V$ is a $\mathcal{N}=1$ vector supermultiplet containing $A^{\mu}$ and $\lambda$, and $\chi$ is a chiral $\mathcal{N}=1$ supermultiplet containing $\lambda'$ and $S'=S+iA^5$. This form of writing the fields follows from the decomposition of the 5D supercharge (which is a Dirac spinor) into two Majorana-type supercharges, which constitute a $\mathcal{N}=2$ superalgebra in 4D. Both $V$ and $\chi$ (and their component fields) are in the adjoint representation of the gauge group $\mathcal{G}$. Using the supermultiplets one can write the original 5D $\mathcal{N}=1$ supersymmetric action Eq.(\ref{eqn:0}) in terms of $\mathcal{N}=1$ 4D superfields and the covariant derivative in the $y$ direction \cite{Hebecker}:
\begin{equation}
S_g = \int\mathrm{d}^5x\mathrm{d}^2\theta\mathrm{d}^2\overline{\theta}\frac{1}{4kg^2}\mathrm{Tr} \left[\frac{1}{4}(W^{\alpha}W_\alpha\delta(\overline{\theta}^2)+\,h.c)+(e^{-2gV}\nabla_ye^{2gV})^2\right] \; , \label{eqn:0a}
\end{equation}
with $W^{\alpha}=-\frac{1}{4}\overline{D}^2e^{-2gV}D_{\alpha} e^{2gV}$. $D_{\alpha}$ is the covariant derivative in the 4D $\mathcal{N}=1$ superspace (see Refs.\cite{Westbook, WessBagger}) and $\nabla_y=\partial_y+\chi$. This action can be expanded and quantised to find the Feynman rules to a given order in the gauge coupling $g$. The details of this procedure can be found in Ref.\cite{Deandrea:2006mh}. Due to the non-renormalisation theorem \cite{NRtheorem}, the beta functions for the couplings of the operators in the superpotential are governed by the wave function renormalisation constants $Z_{ij} = 1 + \delta Z_{ij}$. The Feynman diagrams related to the wave-function renormalisation are given in Fig.\ref{fig:1}.

\par At this point we can note that Higgs superfields and gauge superfields will always propagate into the fifth dimension in our model. However, different possibilities of localisation for the matter superfields can be studied by taking the two limiting cases of superfields containing the SM fermions in the bulk or all superfields containing SM fermions restricted to the brane, respectively. For the case where all fields can propagate in the bulk, the action for the matter fields would be \cite{Deandrea:2006mh}:
\begin{eqnarray}
S_{matter} &=& \int\mbox{d}^8z\mbox{d}y\left\{ \bar{\Phi}_i\Phi_i +
\Phi^c_i\bar{\Phi}^c_i + \Phi^c_i\partial_5\Phi_i \delta(\bar{\theta}) -
\bar{\Phi}_i\partial_5\overline{\Phi}_i^c\delta(\theta) \right. \nonumber \\
& & \left. \hspace{1.5cm} + \tilde{g}(2\bar{\Phi}_iV\Phi_i - 2\Phi_i^cV\bar{\Phi}^c_i +
\Phi^c_i\chi \Phi_i\delta(\bar{\theta}) + \bar{\Phi}_i\bar{\chi}\bar{\Phi}^c_i\delta(\theta)) \right\} \; .
\end{eqnarray}
Again, this action can be expanded and quantised, as detailed in Ref. \cite{Deandrea:2006mh}. Similarly, we can do the same for the case where all superfields containing SM fermions are restricted to the brane. In which case the part of the action involving only gauge and Higgs fields is not modified, whereas the action for the superfields containing the SM fermions becomes 
\begin{equation}
S_{matter} = \int\mbox{d}^8z\mbox{d}y \delta(y) \left\{ \bar{\Phi}_i\Phi_i +
2\tilde{g}\bar{\Phi}_iV\Phi_i \right\} \; .\label{matterbrane}
\end{equation}

\par Before proceeding further we shall recall that the Yukawa couplings in the bulk are forbidden by the 5D $\mathcal{N}=1$ supersymmetry. However, they can be introduced on the branes, which are 4D subspaces with reduced supersymmetry \cite{Deandrea:2006mh}:
\begin{equation}
S_{brane}=\int\mbox{d}^8z\mbox{d}y\delta(y)\left\{\frac{1}{6} \tilde{\lambda}_{ijk}\Phi_i\Phi_j\Phi_k \delta(\bar{\theta}) + \mbox{h.c.} \right\} \;, \label{Sbrane}
\end{equation}

\par Returning to Figs.\ref{fig:1}a)-e) these lead to the usual Minimal Supersymmetric Standard Model (MSSM) if all the extra-dimensional effects were excluded, that is, where the zero modes in Fig.\ref{fig:1}a) and Fig.\ref{fig:1}c) give the beta functions of the usual 4D MSSM. Fig.\ref{fig:1}b) is a new extra-dimensional effect related to the coupling to the chiral superfield originating from the fifth component of the gauge fields, and where the higher KK modes in Figs.\ref{fig:1}a), c), d) and e) are also contributing additional extra-dimensional effects. Note that the mass dimension for the Yukawa couplings in 5D is $-\frac{1}{2}$, therefore, the power law dependence of their evolution, which results from these radiative corrections, becomes significantly important. Also, it should be noted at this point that though the matter fields do not normally receive a wavefunction renormalization in ${\cal N} = 2$ supersymmetry, that is the KK conserved diagrams would not contribute to the Yukawa evolution, ${\cal N} = 2$ supersymmetry has been explicitly broken by the compactification \cite{Deandrea:2006mh}.

\par As such, following the procedures outlined in Ref. \cite{Deandrea:2006mh} the RGEs for the Yukawa couplings in the 5D MSSM, for all three generations propagating in the bulk, is expressed as:
\begin{eqnarray}
16{\pi ^2}\frac{{d{Y_d}}}{{dt}} &=& {Y_d}(3Tr(Y_d^\dag {Y_d}) + Tr(Y_e^\dag {Y_e}) + 3Y_d^\dag {Y_d} + Y_u^\dag {Y_u})\pi S{(t)^2} - {Y_d}\left( \frac{7}{15} g_1^2 + 3 g_2^2 + \frac{16}{3} g_3^2 \right) S(t) \; ,  \nonumber\\
16{\pi ^2}\frac{{d{Y_u}}}{{dt}} &=& {Y_u}(3Tr(Y_u^\dag {Y_u}) + 3Y_u^\dag {Y_u} + Y_d^\dag {Y_d})\pi S{(t)^2} - {Y_u}\left( {\frac{{13}} {{15}}g_1^2 + 3g_2^2 + \frac{{16}}{3}g_3^2} \right)S(t) \; ,  \nonumber\\
16{\pi ^2}\frac{{d{Y_e}}}{{dt}} &=& {Y_e}(3Tr(Y_d^\dag {Y_d}) + Tr(Y_e^\dag {Y_e}) + 3Y_e^\dag {Y_e})\pi S{(t)^2} - {Y_e}\left( {\frac{9} {5}g_1^2 + 3g_2^2} \right)S(t) \; . \label{eqn:912}
\end{eqnarray}
That is, when the energy scale $\mu > 1/R$ or when the energy scale parameter $t= \ln (\mu /{M_Z}) > \ln (\frac{1}{{{M_Z}R}})$ (where we have set ${M_Z}$ as the renormalisation point, and use $S(t) = {e^t}{M_Z}R$) the beta functions become as above, replacing the usual 4D MSSM RGEs. Note that, since we compactify with the symmetry $y \to  - y$ ($S_1/{Z_2}$ symmetry), the integral of the extra-dimension is from $0 \to \pi R$, not from $0 \to 2\pi R$. As such the ${X_\delta }$ parameter is in our case only half of the volume of the $\delta$-dimensional unit sphere (see Ref.\cite{DDG}), i.e. ${X_1} = 1$, and this leads to a factor of two difference (due to the compactification) with respect to other Refs.\cite{DDG,Deandrea:2006mh}.

\par Furthermore, when $0 < t < \ln (\frac{1}{{{M_Z}R}})$ (that is, ${M_Z} < \mu < 1/R$) the Yukawa evolution equations at low energy are dictated by the usual MSSM, where the above equations reduce to the 4D MSSM equations:
\begin{eqnarray}
16{\pi ^2}\frac{{d{Y_d}}}{{dt}} &=& {Y_d}(3Tr(Y_d^\dag {Y_d}) + Tr(Y_e^\dag {Y_e}) + 3Y_d^\dag {Y_d} + Y_u^\dag {Y_u}) - {Y_d} \left( {\frac{7}{{15}}g_1^2 + 3g_2^2 + \frac{{16}}{3}g_3^2} \right) \; ,\nonumber\\
16{\pi ^2}\frac{{d{Y_u}}}{{dt}} &=& {Y_u}(3Tr(Y_u^\dag {Y_u}) + 3Y_u^\dag {Y_u} + Y_d^\dag {Y_d}) - {Y_u}\left( {\frac{{13}}{{15}}g_1^2 + 3g_2^2 + \frac{{16}}{3}g_3^2} \right) \; , \nonumber\\
16{\pi ^2}\frac{{d{Y_e}}}{{dt}} &=& {Y_e}(3Tr(Y_d^\dag {Y_d}) + Tr(Y_e^\dag {Y_e}) + 3Y_e^\dag {Y_e}) - {Y_e}\left( {\frac{9}{5}g_1^2 + 3g_2^2} \right)\; . \label{eqn:3}
\end{eqnarray}
Note also that from the KK number conserved one-loop diagram in Fig.\ref{fig:1}a), at each KK level the new excited states exactly mirror the zero mode ground state, where their contributions to the anomalous dimensions are exactly the same as those in the usual 4-dimensional MSSM. As a result, when the energy $\mu  = 1/R$, or $S(t)=1$, the part in the beta function associated with the gauge fields in Eqs.(\ref{eqn:912}) reduce to the normal 4D MSSM formalism in Eqs.(\ref{eqn:3}).

\par For the second case we shall consider, that of the matter fields localised on the brane, we have (in the Feynman diagrams, only Fig.\ref{fig:1}a), c) and d) will contribute to the wave function of the matter fields, while the non-zero integer numbers inside the bracket labeling the KK states of the gauge and Higgs superfields respectively)
\begin{eqnarray}
16{\pi ^2}\frac{{d{Y_d}}}{{dt}} &=& {Y_d}(3Tr(Y_d^\dag {Y_d}) + Tr(Y_e^\dag {Y_e}) + (6Y_d^\dag {Y_d} + 2Y_u^\dag {Y_u}) S{(t)}) - {Y_d} \left( \frac{19}{30} g_1^2 + \frac{9}{2} g_2^2 + \frac{32}{3} g_3^2 \right) S(t) \; , \nonumber\\
16{\pi ^2}\frac{{d{Y_u}}}{{dt}} &=& {Y_u}(3Tr(Y_u^\dag {Y_u}) + (6Y_u^\dag {Y_u} + 2Y_d^\dag {Y_d}) S{(t)}) - {Y_u}\left( {\frac{{43}}{{30}}g_1^2 + \frac{{9}}{{2}}g_2^2 + \frac{{32}}{3}g_3^2} \right)S(t) \; , \nonumber\\
16{\pi ^2}\frac{{d{Y_e}}}{{dt}} &=& {Y_e}(3Tr(Y_d^\dag {Y_d}) + Tr(Y_e^\dag {Y_e}) + 6Y_e^\dag {Y_e}S{(t)}) - {Y_e}\left( {\frac{33}{10}g_1^2 + \frac{9}{2}g_2^2} \right)S(t) \; . \label{eqn:912b}
\end{eqnarray}

\par As such, if we write the evolution of the gauge couplings in 4-dimensions as:
\begin{eqnarray}
16{\pi ^2}\frac{{d{g_i}}}{{dt}} &=& {b_i}{g_i}^3 \; , \label{eqn:10}
\end{eqnarray}
where in the 4D MSSM the parameters $b_i$ read $({b_1},{b_2},{b_3})=(\frac{{33}}{5},1, - 3)$ \cite{Babu:1987im}, using an $SU(5)$ normalisation. Then if we consider our 5D theory as effective up to a scale $\Lambda$, then between the scale $R^{-1}$ (the compactification scale of the single flat extra-dimension), where the first KK states are excited and the cut-off scale $\Lambda$, for the gauge couplings, there are finite quantum corrections from the $\Lambda R$ number of KK states. As a result, once the KK states are excited, these couplings exhibit power law dependencies on $\Lambda$. This can be illustrated if $\Lambda R \gg 1$, to a very good accuracy, with the generic 4D beta function of the gauge couplings with the power law evolution behaviour \cite{YukawaXD, Cornell:2010sz}
\begin{equation}
\beta^{4D} \to \beta^{4D} + \left( S(\mu) -1 \right) \tilde{\beta} \; ,
\end{equation}
where $\tilde{\beta}$ is a generic contribution from a single KK level, and where its coefficient is not a constant but instead $S(\mu) = \mu R$, with $\mu^{max} = \Lambda$, reflecting the power law running behaviour. Therefore, in terms of the scale parameter $t$, the evolution of the gauge couplings can be written as:
\begin{eqnarray}
16{\pi ^2}\frac{{d{g_i}}}{{dt}} &=& [{b_i} + (S(t) - 1){{\tilde b}_i}]{g_i}^3 \; . \label{eqn:15}
\end{eqnarray}
Next we consider the beta functions of the gauge couplings in the 5D MSSM. In fact, after compactification of the 5D MSSM, we have two 4D $\mathcal{N}=1$ chiral supermultiplets, ${\Phi}$ and ${\Phi ^c}$, where the zero modes of ${\Phi}$ give us the normal matter fields and Higgs fields as well as their super partners, while ${\Phi ^c}$ is a new supermultiplet. In the component field formalism, at each KK level, aside from the quantum corrections that mirror those of the 4D MSSM, the only new one-loop contributions to the ${A_\mu }$ Feynman diagrams are from the wave function renormalisation of ${A_\mu }$ (which contribute via the coupling of ${A_\mu }$ to the complex scalar field and its super-partner in the superfield $\chi$, and the coupling of ${A_\mu }$ with the new fermion field and its super-partner in the superfield ${\Phi ^c}$ associated with the two doublets of the Higgs fields and the matter fields respectively in the bulk). This then gives rise to the master beta functions of the gauge couplings in the 5D MSSM as follows:
\begin{eqnarray}
({\tilde b}_1,{\tilde b}_2, {\tilde b}_3) &=& \left( \frac{6}{5}, - 2, - 6 \right) + 4\eta \; , \label{eqn:26}
\end{eqnarray}
where $\eta$ represents the number of generations of fermions which propagate in the bulk. So in the two cases we shall consider, that of all fields propagating in the bulk, i.e. $\eta = 3$, we have \cite{YukawaXD}:
\begin{eqnarray}
{{\tilde b}_i} &=& \left( \frac{{66}}{5},10,6 \right) \; . \label{eqn:16}
\end{eqnarray}
Similarly, for all our matter fields localised to the 3-brane (that is, $\eta = 0$), we have:
\begin{eqnarray}
{{\tilde b}_i} &=& \left( \frac{{6}}{5},-2,-6 \right) \; . \label{eqn:17}
\end{eqnarray}

\par
\begin{figure}[t]
\begin{center}
\includegraphics[width=0.45\textwidth]{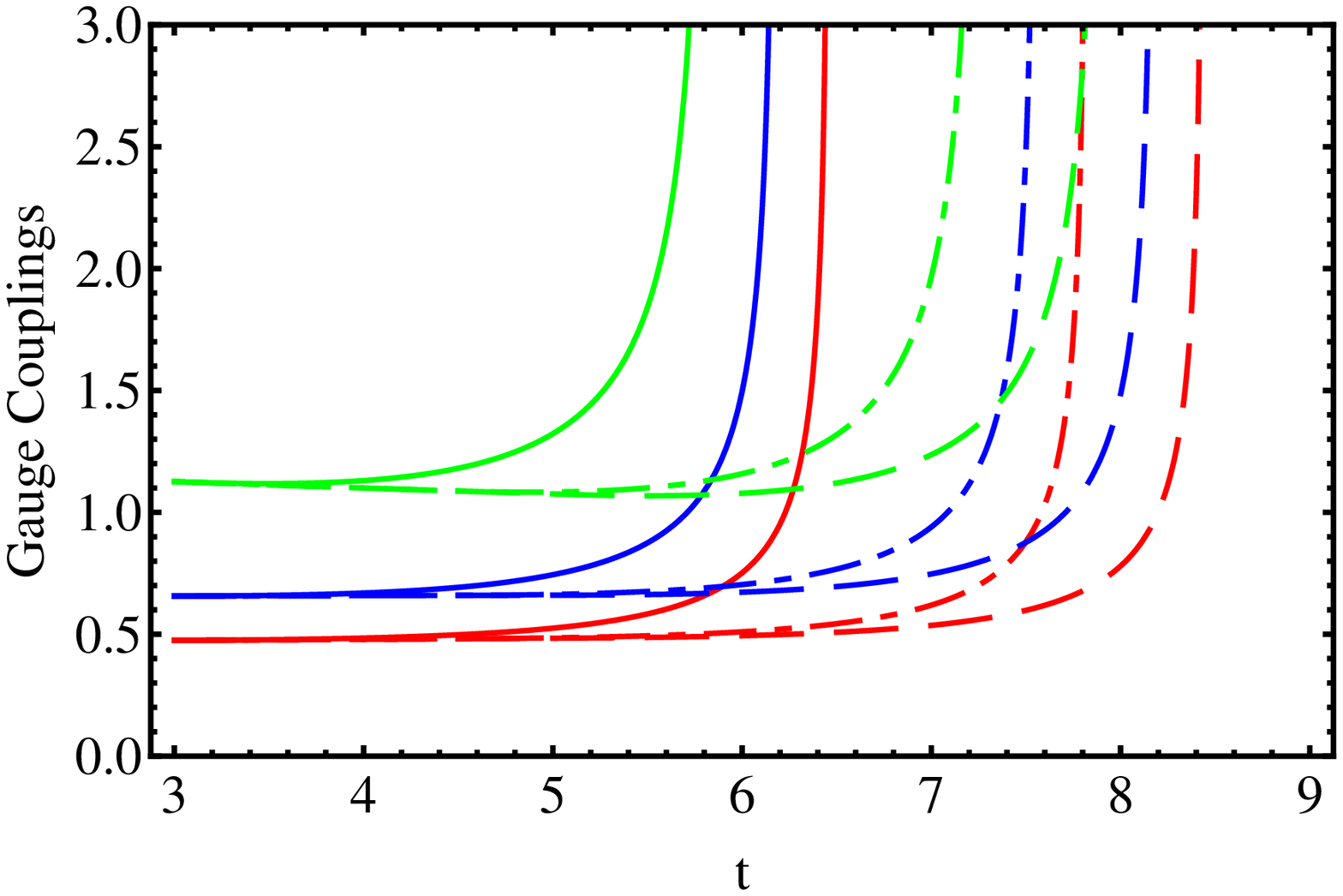}\hspace{0.05\textwidth}
\includegraphics[width=0.45\textwidth]{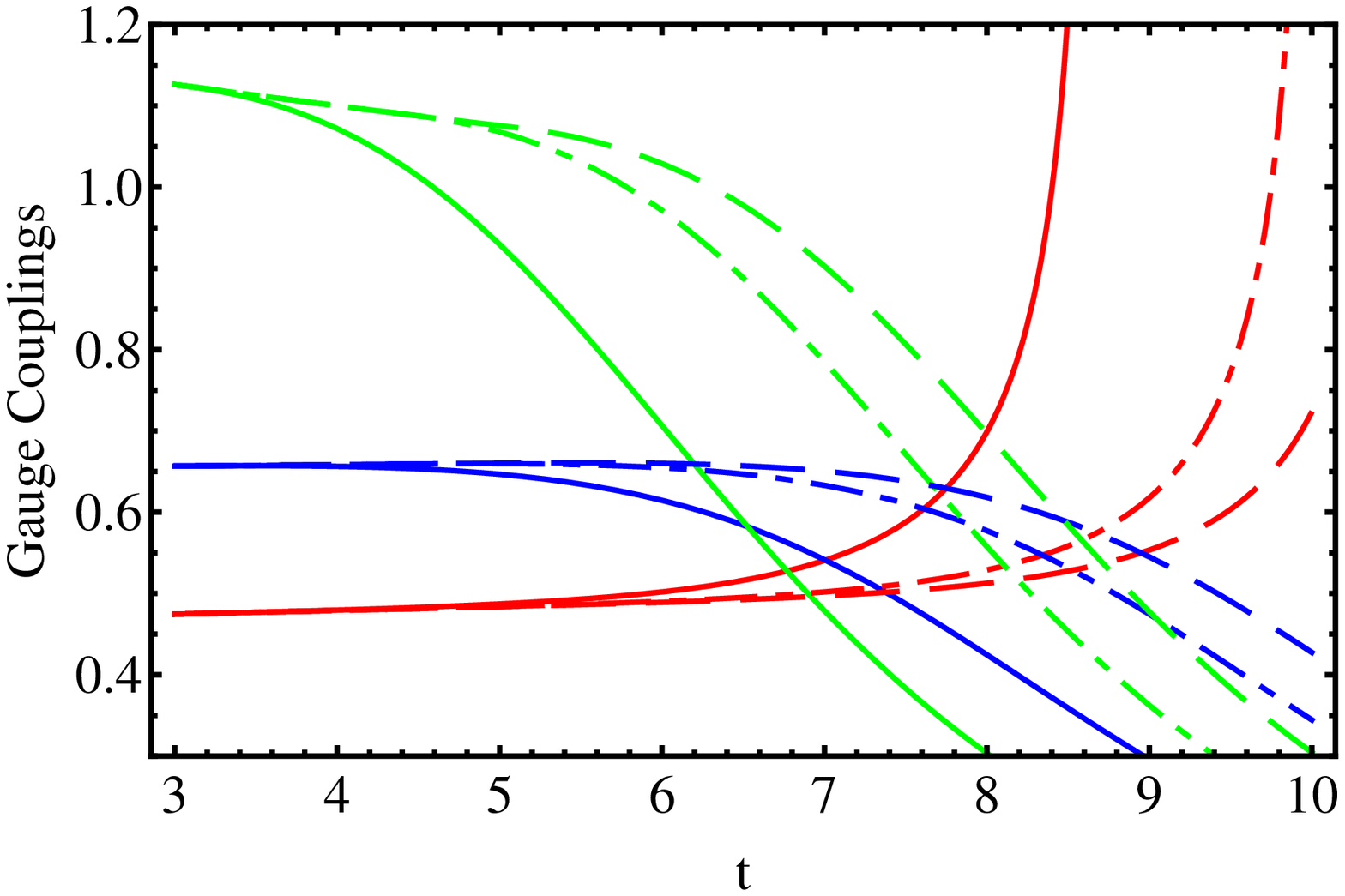}
\caption{\sl (Colour online) Gauge couplings ($g_1$ (red), $g_2$ (blue), $g_3$ (green) with: in the left panel, all matter fields in the bulk; and the right panel for all matter fields on the brane; for three different values of the compactification scales (2 TeV (solid line), 8 TeV (dot-dashed line), 15 TeV (dashed line)) as a function of the scale parameter $t$.}
\label{fig:2}
\end{center}
\end{figure}

 In Fig.\ref{fig:2} we have plotted the evolutions of the brane localised and bulk field cases for several choices of compactification scales for the extra-dimension ($R$). From these plots, and the discussion given in Ref.\cite{YukawaXD}, we find that for the three gauge coupling constants to approach a small region at some value of $t$ requires an extremely large value of $1/R$, which is of no phenomenological interest at present. For the case of our fields being brane localised, the extra-dimensions naturally lead to gauge coupling unification at an intermediate mass scale for the compactification radii considered here.

\section{Scaling of the CKM matrix}\label{sec:3}

\par The RGEs are an important tool for investigating the properties of the quark masses and the CKM matrix at different energy scales, this therefore strongly constrains the possible symmetries or textures at the grand unification scale. Having derived a set of equations for the evolutions of the Yukawa and gauge couplings in the previous section, we now develop the evolutions of the CKM matrix in the 5D MSSM for our two cases. Consider first the case of all the matter fields propagating in the bulk. From the Yukawa evolution equations Eqs.(\ref{eqn:912}) we can convert them to the following form:
\begin{eqnarray}
16{\pi ^2}\frac{{d{Y_d}}}{{dt}} &=& {Y_d}\left\{ {{T_d}\pi {S^2} - {G_d} + (3Y_d^\dag {Y_d} + Y_u^\dag {Y_u})\pi {S^2}} \right\} \; , \nonumber\\
16{\pi ^2}\frac{{d{Y_u}}}{{dt}} &=& {Y_u}\left\{ {{T_u}\pi {S^2} - {G_u} + (3Y_u^\dag {Y_u} + Y_d^\dag {Y_d})\pi {S^2}} \right\} \; , \nonumber\\
16{\pi ^2}\frac{{d{Y_e}}}{{dt}} &=& {Y_e}\left\{ {{T_e}\pi {S^2} - {G_e} + (3Y_e^\dag {Y_e})\pi {S^2}} \right\} \; ,  \label{eqn:32}
\end{eqnarray}
where ${T_d}=3Tr(Y_d^\dag {Y_d}) + Tr(Y_e^\dag {Y_e})$, ${G_d}=\left( {\frac{7}{{15}}g_1^2 + 3g_2^2 + \frac{{16}}{3}g_3^2} \right)S(t)$, ${T_u}=3Tr(Y_u^\dag {Y_u})$,${G_u}=\left( {\frac{{13}}{{15}}g_1^2 + 3g_2^2 + \frac{{16}}{3}g_3^2} \right)S(t)$, ${T_e}=3Tr(Y_d^\dag {Y_d}) + Tr(Y_e^\dag {Y_e})$ and ${G_e}=\left( {\frac{9}{5}g_1^2 + 3g_2^2} \right)S(t)$. Further, the evolution of the square of the Yukawa coupling matrices becomes:
\begin{eqnarray}
16{\pi ^2}\frac{{d(Y_d^\dag {Y_d})}}{{dt}} &=& 2({T_d}\pi {S^2} - {G_d})(Y_d^\dag {Y_d}) + 6{(Y_d^\dag {Y_d})^2}\pi {S^2} + [(Y_u^\dag {Y_u})(Y_d^\dag {Y_d}) + (Y_d^\dag {Y_d})(Y_u^\dag {Y_u})]\pi {S^2}\; ,  \nonumber\\
16{\pi ^2}\frac{{d(Y_u^\dag {Y_u})}}{{dt}} &=& 2({T_u}\pi {S^2} - {G_u})(Y_u^\dag {Y_u}) + 6{(Y_u^\dag {Y_u})^2}\pi {S^2} + [(Y_u^\dag {Y_u})(Y_d^\dag {Y_d}) + (Y_d^\dag {Y_d})(Y_u^\dag {Y_u})]\pi {S^2}\; ,  \nonumber\\
16{\pi ^2}\frac{{d(Y_e^\dag {Y_e})}}{{dt}} &=& 2({T_e}\pi {S^2} - {G_e})(Y_e^\dag {Y_e}) + 6{(Y_e^\dag {Y_e})^2}\pi {S^2}\; . \label{eqn:34}
\end{eqnarray}

\par The square of the quark Yukawa coupling matrices can be diagonalized by using two unitary matrices $U$ and $V$,
\begin{eqnarray}
\mathrm{diag}\left(f_u^2,f_c^2,f_t^2\right)&=& U Y_u^\dag {Y_u}{U^\dag }\; ,  \nonumber\\
\mathrm{diag}\left(h_d^2,h_s^2,h_b^2\right)&=& V Y_d^\dag {Y_d}{V^\dag }\; , \label{eqn:341}
\end{eqnarray}
in which $f_u^2$, $f_c^2$, $f_t^2$ and $h_d^2$, $h_s^2$, $h_b^2$ are the eigenvalues of $Y_u^\dag Y_u$ and $ Y_d^\dag Y_d$ respectively. It follows that the CKM matrix appears as a result of the transition from the quark flavor eigenstates to the quark mass eigenstates upon this diagonalization of the quark mass matrices:
\begin{eqnarray}
{V_{CKM}} = U{V^\dag }\; . \label{eqn:342}
\end{eqnarray}
By imposing the unitary transformation Eq.(\ref{eqn:341}), on both sides of the evolution equations of $Y_u^\dag Y_u$ and $Y_d^\dag Y_d$, and taking the diagonal elements, we obtain the following two relations:
\begin{eqnarray}
16{\pi ^2}\frac{{df_i^2}}{{dt}} &=& f_i^2[2({T_u}\pi {S^2} - {G_u}) + 6\pi {S^2}f_i^2 + 2\pi {S^2}\sum\limits_j  {h_j ^2} {\left| {{V_{ij}}} \right|^2}]\;  \nonumber\\
16{\pi ^2}\frac{{dh_j ^2}}{{dt}} &=& h_j ^2[2({T_d}\pi {S^2} - {G_d}) + 6\pi {S^2}h_j ^2 + 2\pi {S^2}\sum\limits_i {f_i^2} {\left| {{V_{ij}}} \right|^2}] \;, \label{eqn:36}
\end{eqnarray}
where $V_{ij}$ are the elements of $V_{CKM}$. These along with the following equation
\begin{eqnarray}
16{\pi ^2}\frac{{dy_e^2}}{{dt}} &= &y_e^2[2({T_e}\pi {S^2} - {G_e}) + 6\pi {S^2}y_e^2] \; , \label{eqn:361}
\end{eqnarray}
for the lepton sector, constitute a complete set of coupled differential equations for the three families, in which $y_e^2=\mathrm{diag}\left(y_e^2,y_{\mu}^2,y_{\tau}^2\right)=Y_e^\dag {Y_e}$.

\par Considering the variation of the square of the quark Yukawa couplings, we may impose two new unitary matrices to make them diagonal. Thus, by applying Eq.(\ref{eqn:342}), we are led to the variation of the CKM matrix and thus its evolution equation when the energy scale is beyond the threshold $R^{-1}$:
\begin{eqnarray}
16{\pi ^2}\frac{{d{V_{ik}}}}{{dt}} = \pi {S^2}\left[\sum\limits_{m,j \ne i} {\frac{{f_i^2 + f_j^2}}{{f_i^2 - f_j^2}}} h_m^2{V_{im}}V_{jm}^*{V_{jk}} + \sum\limits_{j,m  \ne k} {\frac{{h_k^2 + h_m^2}}{{h_k^2 - h_m^2}}} f_j^2V_{jm}^*{V_{jk}}{V_{im}}\right] \; . \label{eqn:37}
\end{eqnarray}
\par Likewise, for the case where the matter fields are localised to the brane, Eq.(\ref{eqn:912b}) gives us the following:
\begin{eqnarray}
16{\pi ^2}\frac{{d{Y_d}}}{{dt}} &=& {Y_d}\left\{ {{T_d} - {G_{d}} + (6Y_d^\dag {Y_d} + 2Y_u^\dag {Y_u})S} \right\}\; , \nonumber\\
16{\pi ^2}\frac{{d{Y_u}}}{{dt}} &=& {Y_u}\left\{ {{T_u} - {G_{u}} + (6Y_u^\dag {Y_u} + 2Y_d^\dag {Y_d})S} \right\}\; , \nonumber\\
16{\pi ^2}\frac{{d{Y_e}}}{{dt}} &=& {Y_e}\left\{ {{T_e} - {G_{e}} + (6Y_e^\dag {Y_e})S} \right\}\; , \label{eqn:42}
\end{eqnarray}
where ${T_d}=3Tr(Y_d^\dag {Y_d}) + Tr(Y_e^\dag {Y_e})$, ${G_{d}}=\left( {\frac{19}{{30}}g_1^2 + \frac{9}{{2}}g_2^2 + \frac{{32}}{3}g_3^2} \right)S(t)$, ${T_u}=3TrY_u^\dag {Y_u}$, ${G_{u}}=\left( {\frac{{43}}{{30}}g_1^2 + \frac{9}{{2}}g_2^2 + \frac{{32}}{3}g_3^2} \right)S(t)$, ${T_e}=3Tr(Y_d^\dag {Y_d}) + Tr(Y_e^\dag {Y_e})$ and ${G_e}= \left( {\frac{33}{10}g_1^2 +\frac{9}{{2}}g_2^2} \right)S(t)$. As was done earlier, these can be manipulated to:
\begin{eqnarray}
16{\pi ^2}\frac{{d(Y_d^\dag {Y_d})}}{{dt}} &=& 2({T_d} - {G_{d}})(Y_d^\dag {Y_d}) + 12{(Y_d^\dag {Y_d})^2}S + 2[(Y_u^\dag {Y_u})(Y_d^\dag {Y_d}) + (Y_d^\dag {Y_d})(Y_u^\dag {Y_u})]S \; , \nonumber \\
16{\pi ^2}\frac{{d(Y_u^\dag {Y_u})}}{{dt}} &=& 2({T_u} - {G_{u}})(Y_u^\dag {Y_u}) + 12{(Y_u^\dag {Y_u})^2}S + 2[(Y_u^\dag {Y_u})(Y_d^\dag {Y_d}) + (Y_d^\dag {Y_d})(Y_u^\dag {Y_u})]S \; , \nonumber \\
16{\pi ^2}\frac{{d(Y_e^\dag {Y_e})}}{{dt}} &= &2({T_e} - {G_{e}})(Y_e^\dag {Y_e}) + 12{(Y_e^\dag {Y_e})^2}S \; . \label{eqn:44}
\end{eqnarray}
Giving us
\begin{eqnarray}
16{\pi ^2}\frac{{df_i^2}}{{dt}} &=& f_i^2[2({T_u} - {G_{u}}) + 12 S f_i^2 + 4 S\sum\limits_j  {h_j ^2} {\left| {{V_{ij}}} \right|^2}] \; , \nonumber\\
16{\pi ^2}\frac{{dh_j^2}}{{dt}} &=& h_j^2[2({T_d} - {G_{d}}) + 12 S h_j^2 + 4 S\sum\limits_i {f_i^2} {\left| {{V_{ij}}} \right|^2}] \; , \nonumber \\
16{\pi ^2}\frac{{dy_e^2}}{{dt}} &=& y_e^2[2({T_e} - {G_{e}}) + 12 S y_e^2] \; , \label{eqn:46}
\end{eqnarray}
as well as the evolution equation of the CKM matrix elements
\begin{eqnarray}
16{\pi ^2}\frac{{d{V_{ik}}}}{{dt}} = 2{S}\left[\sum\limits_{m,j \ne i} {\frac{{f_i^2 + f_j^2}}{{f_i^2 - f_j^2}}} h_m^2{V_{im}}V_{jm}^*{V_{jk}} + \sum\limits_{j,m  \ne k} {\frac{{h_k^2 + h_m^2}}{{h_k^2 - h_m^2}}} f_j^2V_{jm}^*{V_{jk}}{V_{im}}\right] \; , \label{eqn:47}
\end{eqnarray}
in which the $f_i^2,h_j^2,y_e^2$ are the eigenvalues of $Y_u^\dag {Y_u},Y_d^\dag {Y_d},Y_e^\dag {Y_e}$ as defined earlier.

\par However, when the energy is below the compactification scale, we have the usual 4D MSSM evolution equations for the Yukawa couplings and the CKM matrix
\begin{eqnarray}
16{\pi ^2}\frac{{df_i^2}}{{dt}} &=& f_i^2[2({T_u} - {G_{u}}) + 6f_i^2 + 2\sum\limits_j {h_j ^2} {\left| {{V_{ij}}} \right|^2}] \; , \nonumber\\
16{\pi ^2}\frac{{dh_j^2}}{{dt}} &=& h_j^2[2({T_d} - {G_{d}}) + 6h_j^2 + 2\sum\limits_i {f_i^2} {\left| {{V_{ij}}} \right|^2}] \; , \nonumber \\
16{\pi ^2}\frac{{dy_e^2}}{{dt}} &=& y_e^2[2({T_e} - {G_{e}}) + 6y_e^2] \; , \label{eqn:471}
\end{eqnarray}
and
\begin{eqnarray}
16{\pi ^2}\frac{{d{V_{ik}}}}{{dt}} = \left[\sum\limits_{m,j \ne i} {\frac{{f_i^2 + f_j^2}}{{f_i^2 - f_j^2}}} h_m^2{V_{im}}V_{jm}^*{V_{jk}} + \sum\limits_{j,m  \ne k} {\frac{{h_k^2 + h_m^2}}{{h_k^2 - h_m^2}}} f_j^2V_{jm}^*{V_{jk}}{V_{im}}\right] \; ,\label{eqn:472}
\end{eqnarray}
where ${T_u}=3Tr(Y_u^\dag {Y_u})$, ${G_{u}}={\frac{{13}}{{15}}g_1^2 + 3g_2^2 + \frac{{16}}{3}g_3^2}$, ${T_d}=3Tr(Y_d^\dag {Y_d}) + Tr(Y_e^\dag {Y_e})$, ${G_{d}}={\frac{7}{{15}}g_1^2 + 3g_2^2 + \frac{{16}}{3}g_3^2}$, ${T_e}=3Tr(Y_d^\dag {Y_d}) + Tr(Y_e^\dag {Y_e})$ and ${G_e}= {\frac{9}{5}g_1^2 +3g_2^2}$.

\par Before proceeding to numerically solve these systems of equations, we choose the initial input values for the gauge couplings, fermion masses and CKM elements at the $M_Z$ scale (as in Ref.\cite{Cornell:2010sz}). The 4D MSSM contains the particle spectrum of a two-Higgs doublet model extension of the SM and the corresponding supersymmetric partners. After the spontaneous breaking of the electroweak symmetry, five physical Higgs particles are left in the spectrum. The two Higgs doublets $H_u$ and $H_d$, with opposite hypercharges, are responsible for the generation of the up-type and down-type quarks respectively. The vacuum expectation values of the neutral components of the two Higgs fields satisfy the relation ${v_u}^2 + {v_d}^2 = {\left( {\frac{{246}}{{\sqrt 2 }}} \right)^2} = {\left( {174GeV} \right)^2}$. The fermion mass matrices appear after the spontaneous symmetry breaking from the fermion-Higgs Yukawa couplings. As a result, the initial Yukawa couplings are given by the ratios of the fermion masses to the appropriate Higgs vacuum expectation values as follows:
\begin{eqnarray}
{f_{u,c,t}} = \frac{{{m_{u,c,t}}}}{{{v_u}}}\;\; , \;\;
{h_{d,s,b}} = \frac{{{m_{d,s,b}}}}{{{v_d}}}\;\; , \;\;
{y_{e,\mu ,\tau }} = \frac{{{m_{e,\mu ,\tau }}}}{{{v_d}}} \; , \label{eqn:48}
\end{eqnarray}
where we define $\tan \beta  = v_u/v_d$, which is the ratio of vacuum expectation values of the two Higgs fields $H_u$ and $H_d$.


\section{Discussion of the results}\label{sec:4}

\par In the previous section we have derived the full set of one-loop coupled RGEs for the Yukawa and gauge couplings, together with the CKM matrix elements for both the universal 5D MSSM and brane localised matter field scenarios. From these complete sets of the RGEs we can obtain the renormalisation group flow of all observables related to up- and down-quark masses and the quark flavor mixings. For our numerical analysis we assume the fundamental scale is not far from the range of LHC scale, and set the compactification scale to be $R^{-1} = 2$ TeV, 8 TeV, and 15 TeV respectively.

\begin{figure}[t]
\begin{center}
\includegraphics[width=0.45\textwidth]{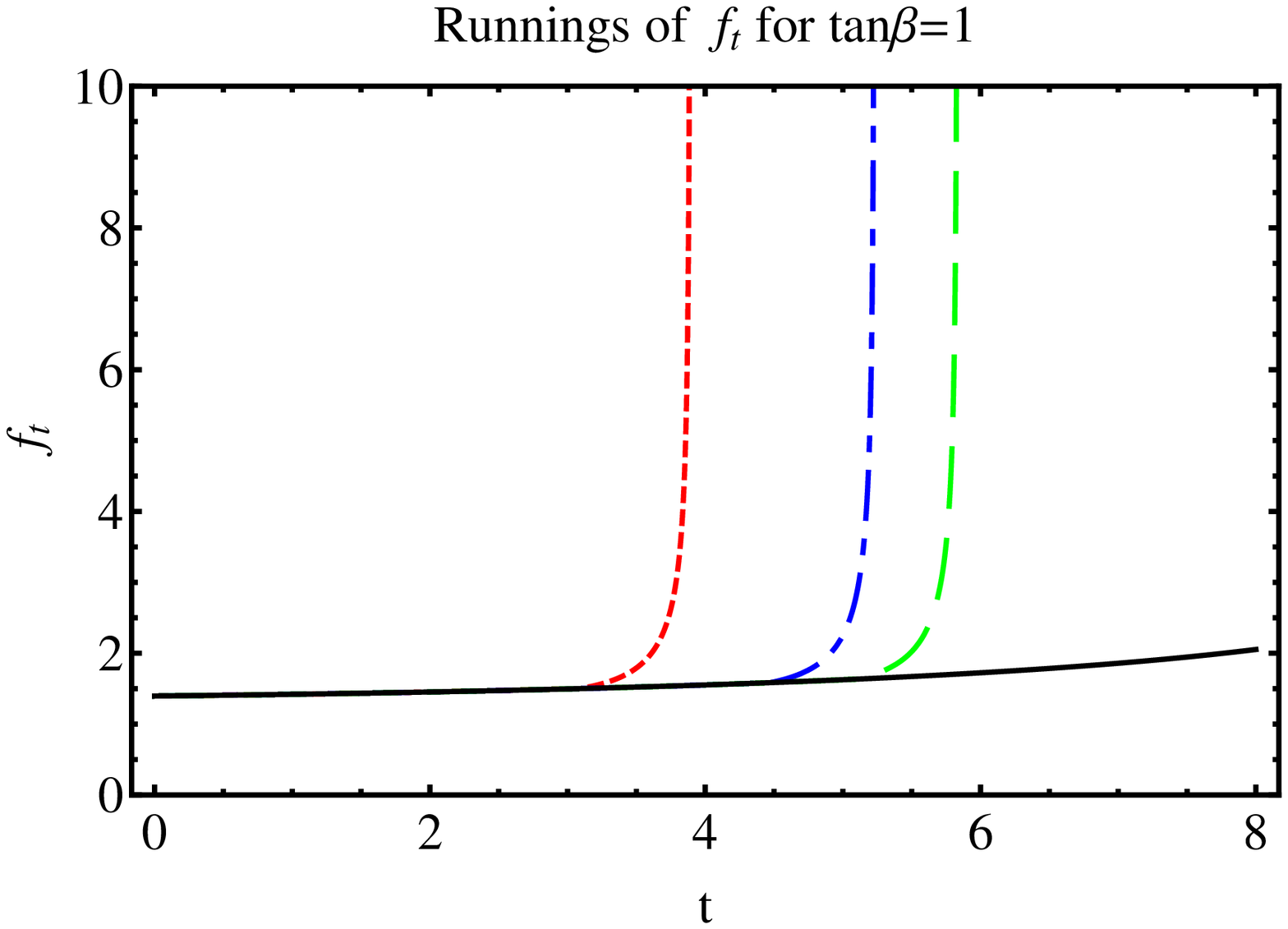}\hspace{0.05\textwidth}
\includegraphics[width=0.45\textwidth]{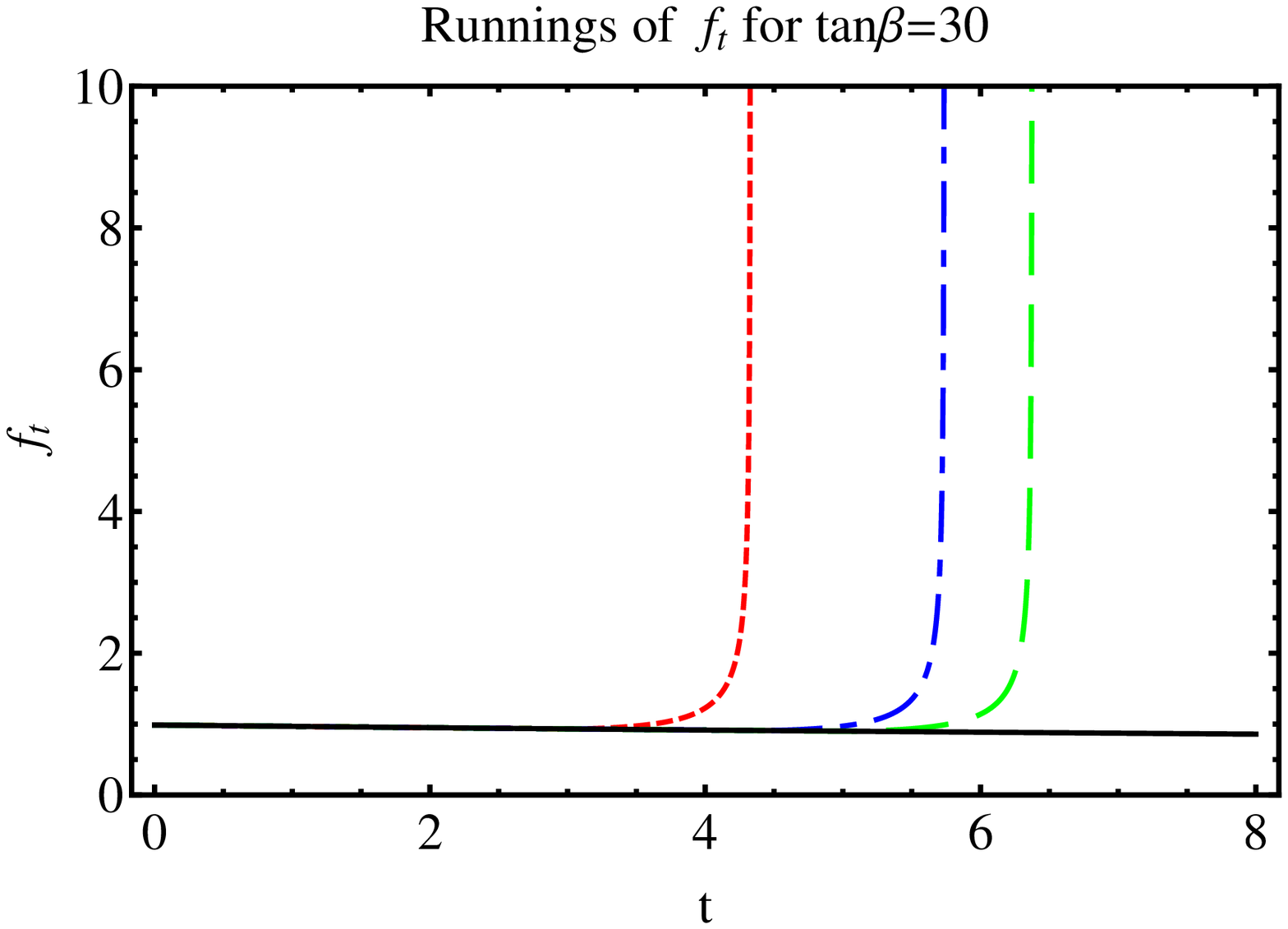}
\caption{\sl (Colour online) The Yukawa coupling $f_t$ for the top quark (in the bulk) as a function of the scale parameter $t$, for (left panel) $\tan\beta=1$ and (right panel) $\tan\beta = 30$ for different compactification scales: $R^{-1}$ = 2 TeV (red, dotted line), 8 TeV (blue, dot-dashed line), and 15 TeV (green, dashed line).}
\label{fig:3}
\end{center}
\end{figure}
\begin{figure}[ht]
\begin{center}
\includegraphics[width=0.45\textwidth]{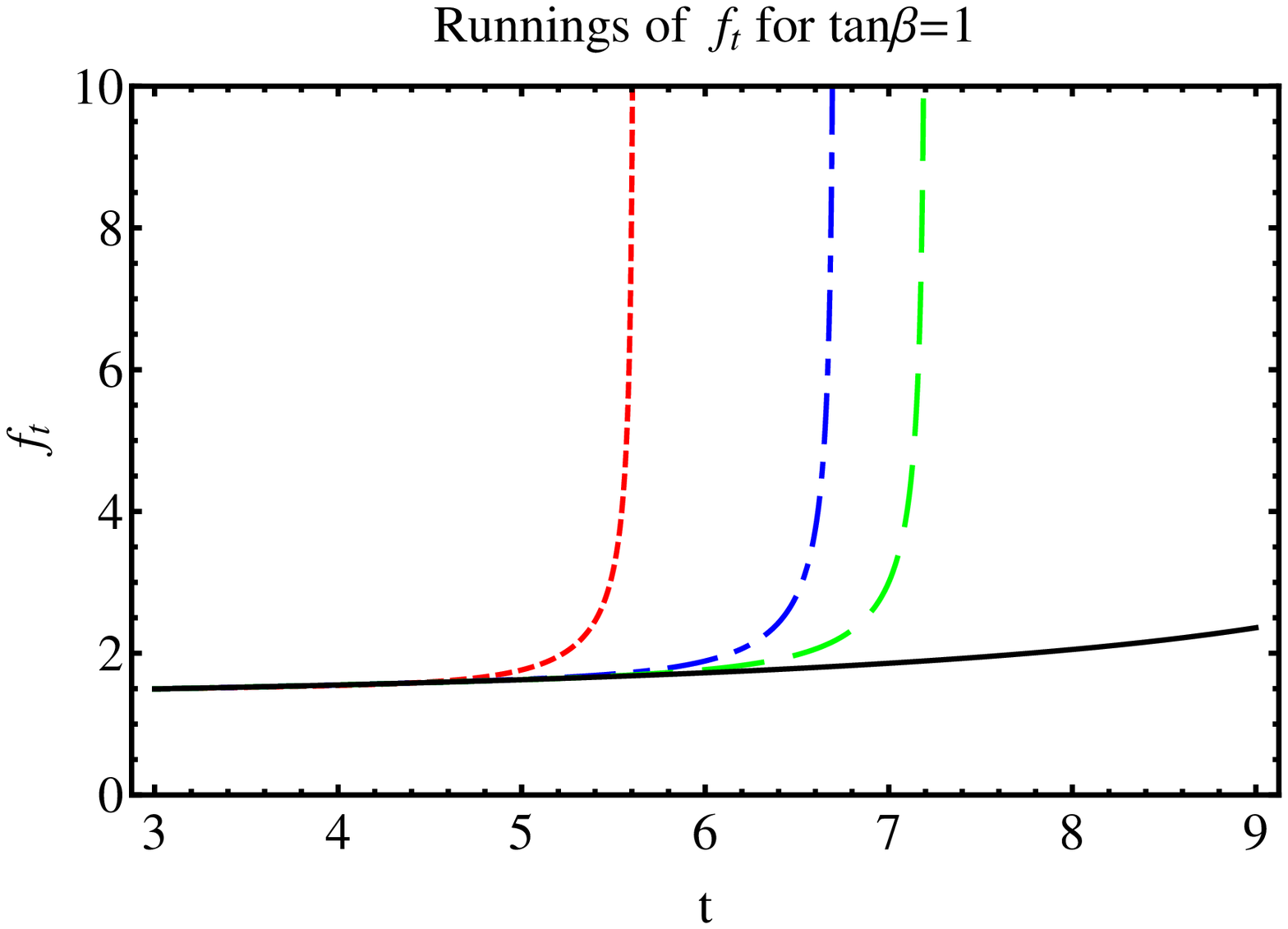}\hspace{0.05\textwidth}
\includegraphics[width= 0.45\textwidth]{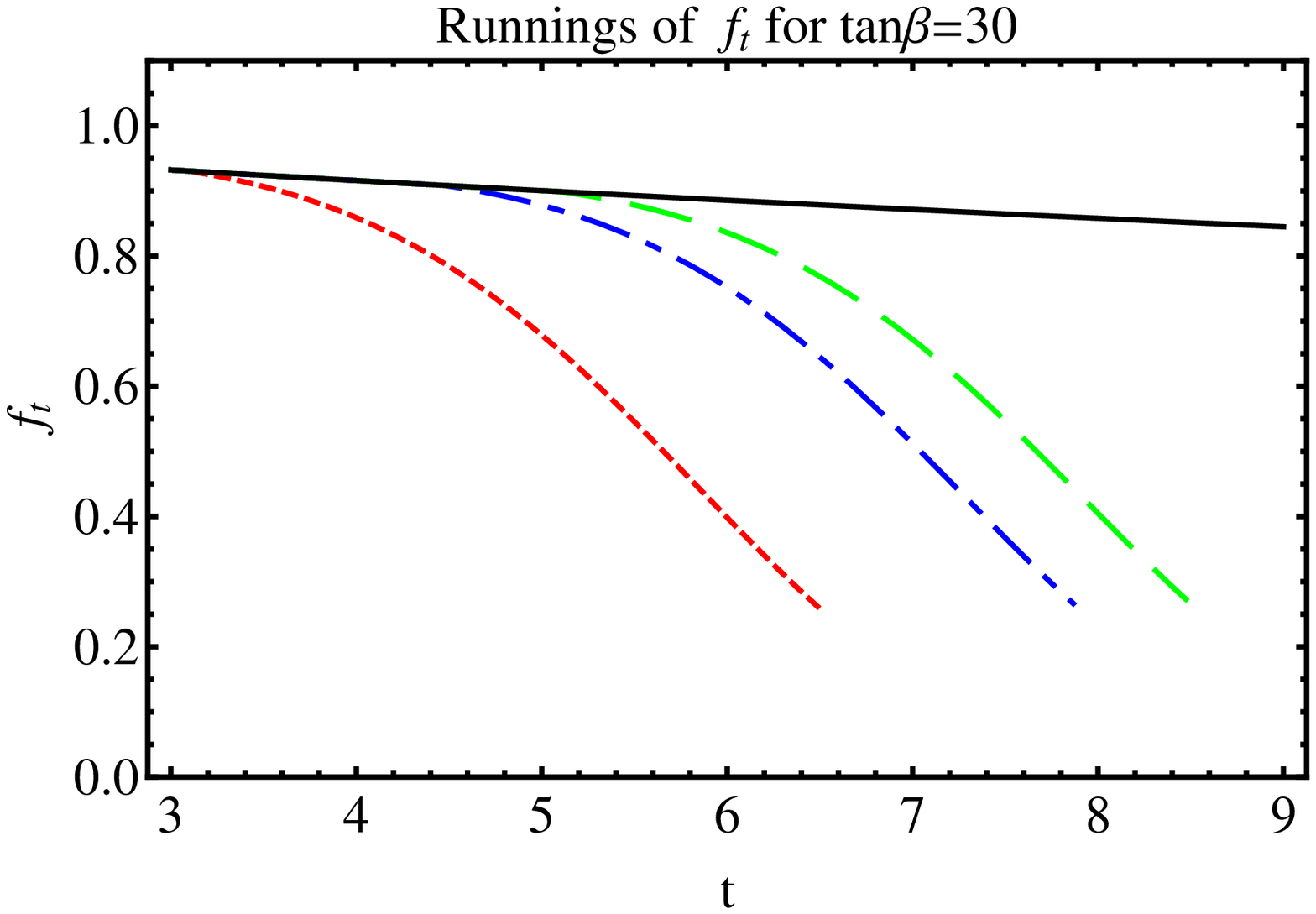}
\caption{\sl (Colour online) The Yukawa coupling $f_t$ for the top quark (on the brane) as a function of the scale parameter $t$, for (left panel) $\tan\beta=1$ and (right panel) $\tan\beta = 30$ for different compactification scales: $R^{-1}$ = 2 TeV (red, dotted line), 8 TeV (blue, dot-dashed line), and 15 TeV (green, dashed line).}
\label{fig:4}
\end{center}
\end{figure}

\par Actually, below the supersymmetric breaking scale the Yukawa and gauge couplings run in the usual logarithmic fashion, giving a rather slow change for their values. Therefore, for supersymmetric breaking theories around TeV scales, for simplicity, we take the supersymmetric breaking scale $M_{SUSY}=M_Z$ in the present numerical study, and run the RGEs from $M_Z$ up to the high energy scales for our three different compactification scales. In Fig.\ref{fig:2} we find the evolution of the gauge couplings have different properties for these two 5D MSSM scenarios.

\par Additionally, as illustrated in Fig.\ref{fig:3}, for the case of universal 5D MSSM, once the first KK threshold is crossed at $\mu = R^{-1}$, the power law running of the various beta functions causes the Yukawa coupling to rapidly increase following the rapid increase in the gauge coupling constants in the left panel of Fig.\ref{fig:2}. From Eq.(\ref{eqn:36}) we can find the quadratic term of $S(t)$ providing a positive contribution to the Yukawa beta functions, which is in contrast to beta functions of the gauge couplings (which include terms only linear in $S(t)$). Therefore, from Eq.(\ref{eqn:36}), the positive contribution from $S(t)$ terms will dominate the negative contributions from the gauge couplings, and cause the Yukawa couplings to increase rapidly. This behaviour can be observed for both small and large $\tan\beta$ cases. However, as illustrated in the first graph of Fig.\ref{fig:3}, for small $\tan\beta$, the Yukawa coupling has a large initial value, therefore it blows up at a relatively low energy as compared with the case for large $\tan\beta$. As a result, as one evolves upward in the scale, the top Yukawa coupling is rising with a fast rate and is pushed up against the Landau pole where it becomes divergent and blows up. In the vicinity of this singular point the perturbative calculation becomes invalid, and the
higher order corrections become significant. The Landau pole also indicates that there is an upper limit on the value of the gauge couplings where new physics must emerge before the Yukawa couplings diverge.

\par In the brane localised matter field scenario, the beta function has only linear terms in $S(t)$, which is comparable with the $S(t)$ term in the beta function for the gauge couplings. As depicted in Fig.\ref{fig:4}, for a small value of $\tan\beta$, we have a large initial value of $f_t$ and the gauge coupling contribution to the Yukawa beta function is sub-dominant only. Therefore, as discussed previously, the Yukawa coupling $f_t$ increases rapidly as one crosses the KK threshold at $\mu = R^{-1}$, resulting in a rapid approach of the singularity before the unification scale is reached. However, for an intermediate value of $\tan\beta$, we have a relative smaller initial condition for the top Yukawa coupling and the Yukawa terms in the beta function become less important. The contributions from the gauge couplings may then become significant, which leads to a net negative contribution to the beta functions. Therefore, the curvature of the trajectory of the top Yukawa evolution might change direction, and the Yukawa evolution will decrease instead of increasing. This behaviour would become more obvious for a large value of $\tan\beta$. As observed in Fig.\ref{fig:4}, for $\tan\beta =30$, we indeed observe the decreasing behaviour of the top Yukawa couplings. This behaviour provides a very clear phenomenological signature, especially for scenarios with a larger $\tan\beta$ and that are valid up to the unification scale where the gauge couplings converge.

\par
\begin{figure}[t]
\begin{center}
\includegraphics[width=0.45\textwidth]{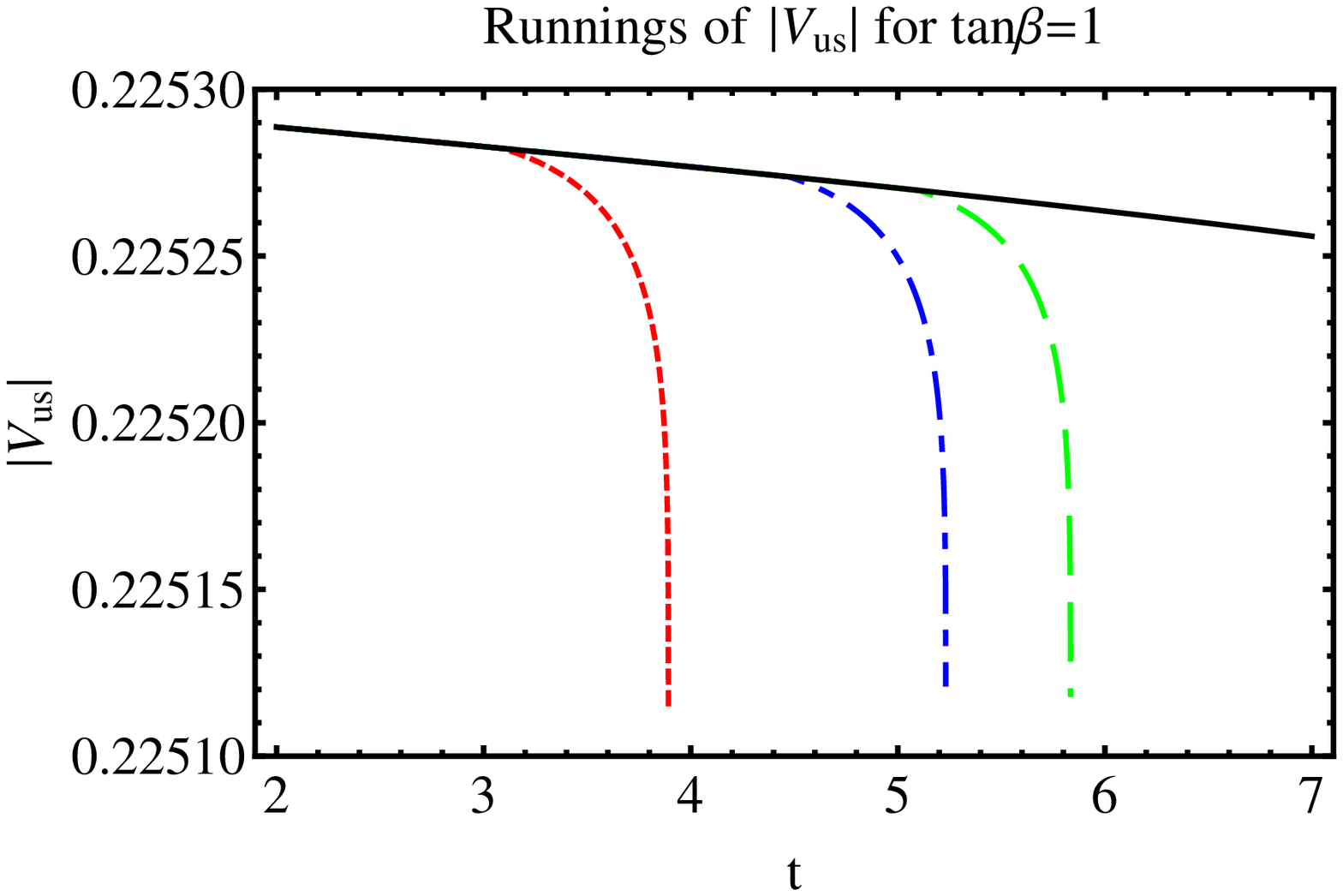}\hspace{0.05\textwidth}
\includegraphics[width=0.45\textwidth]{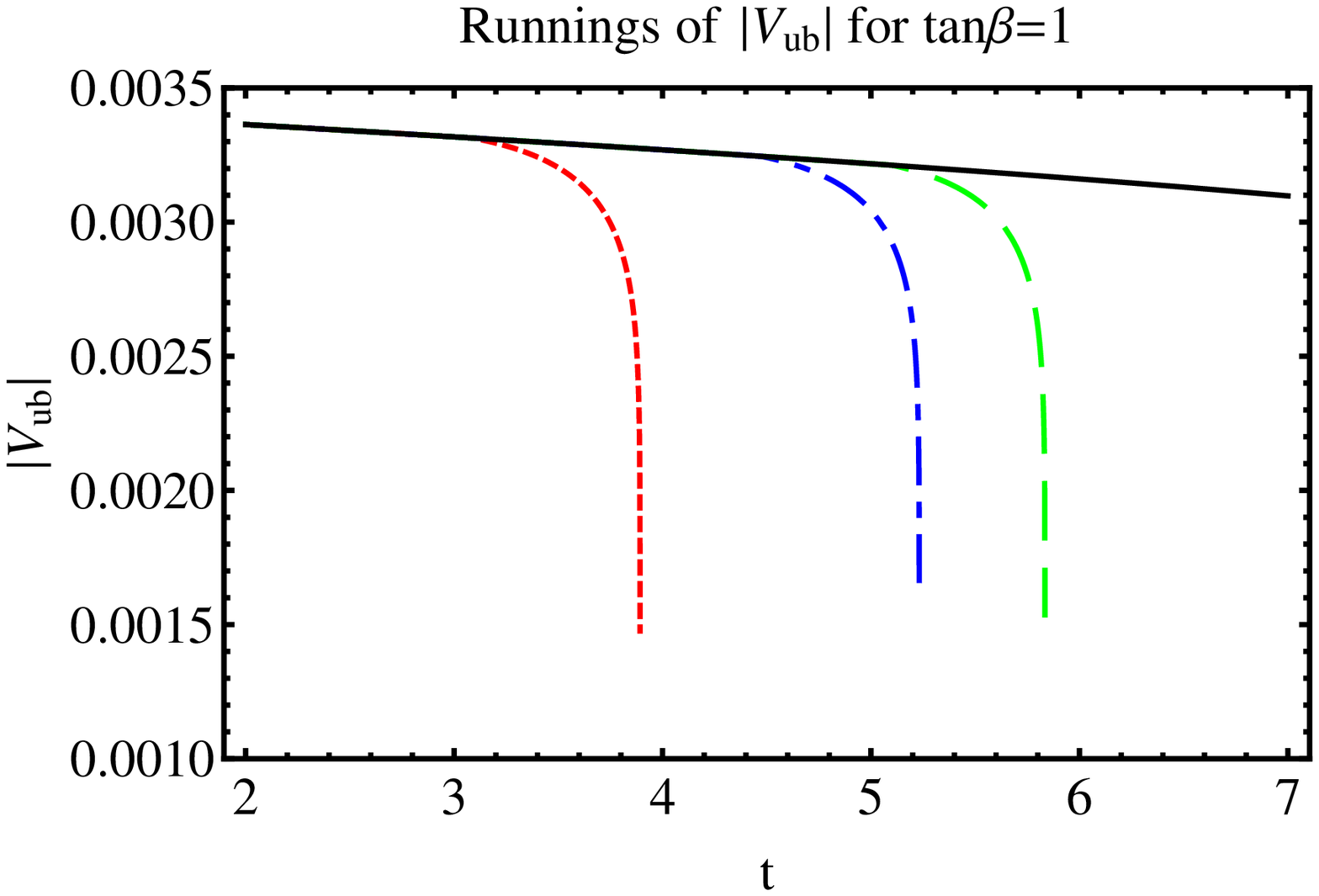}
\includegraphics[width=0.45\textwidth]{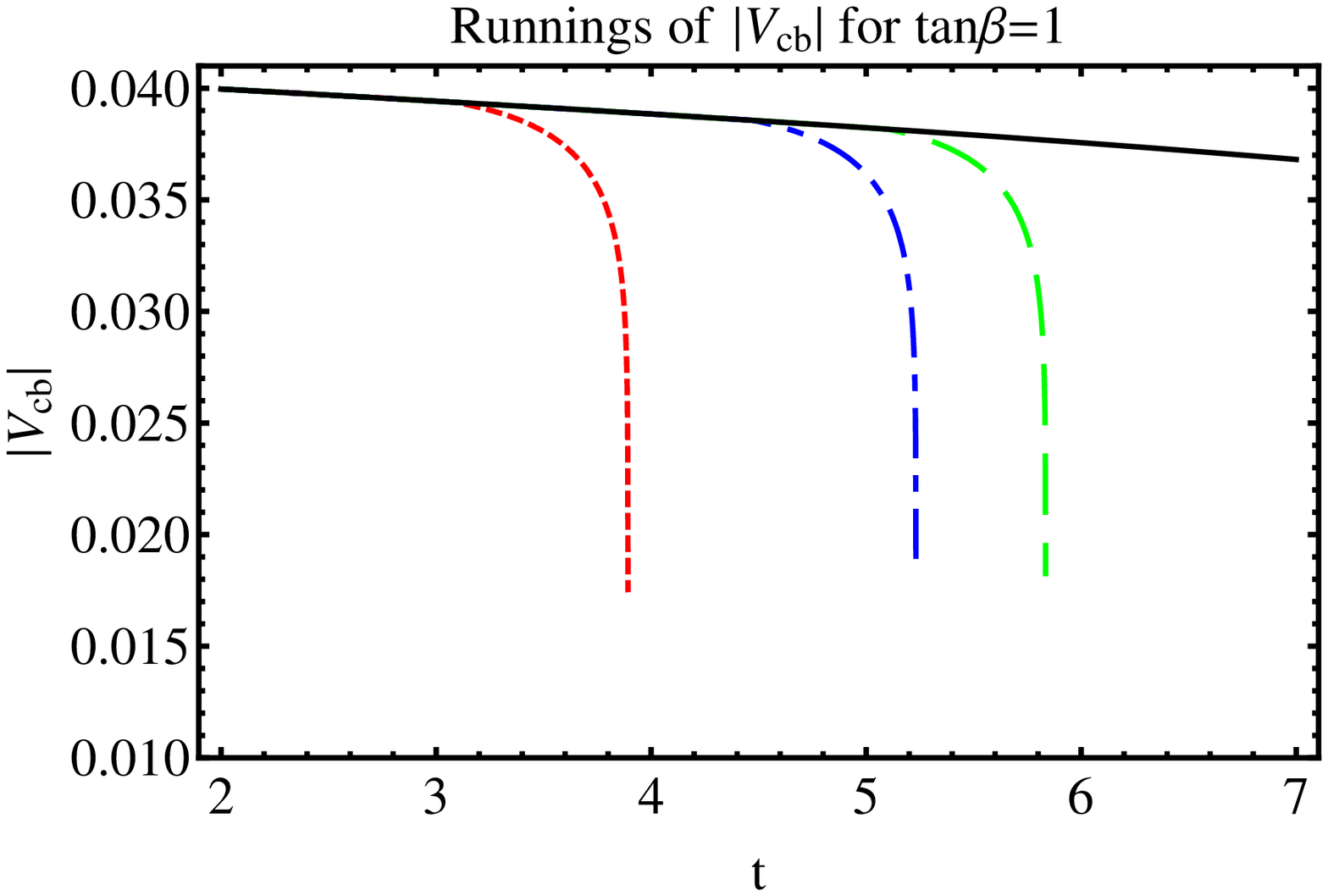}\hspace{0.05\textwidth}
\includegraphics[width=0.45\textwidth]{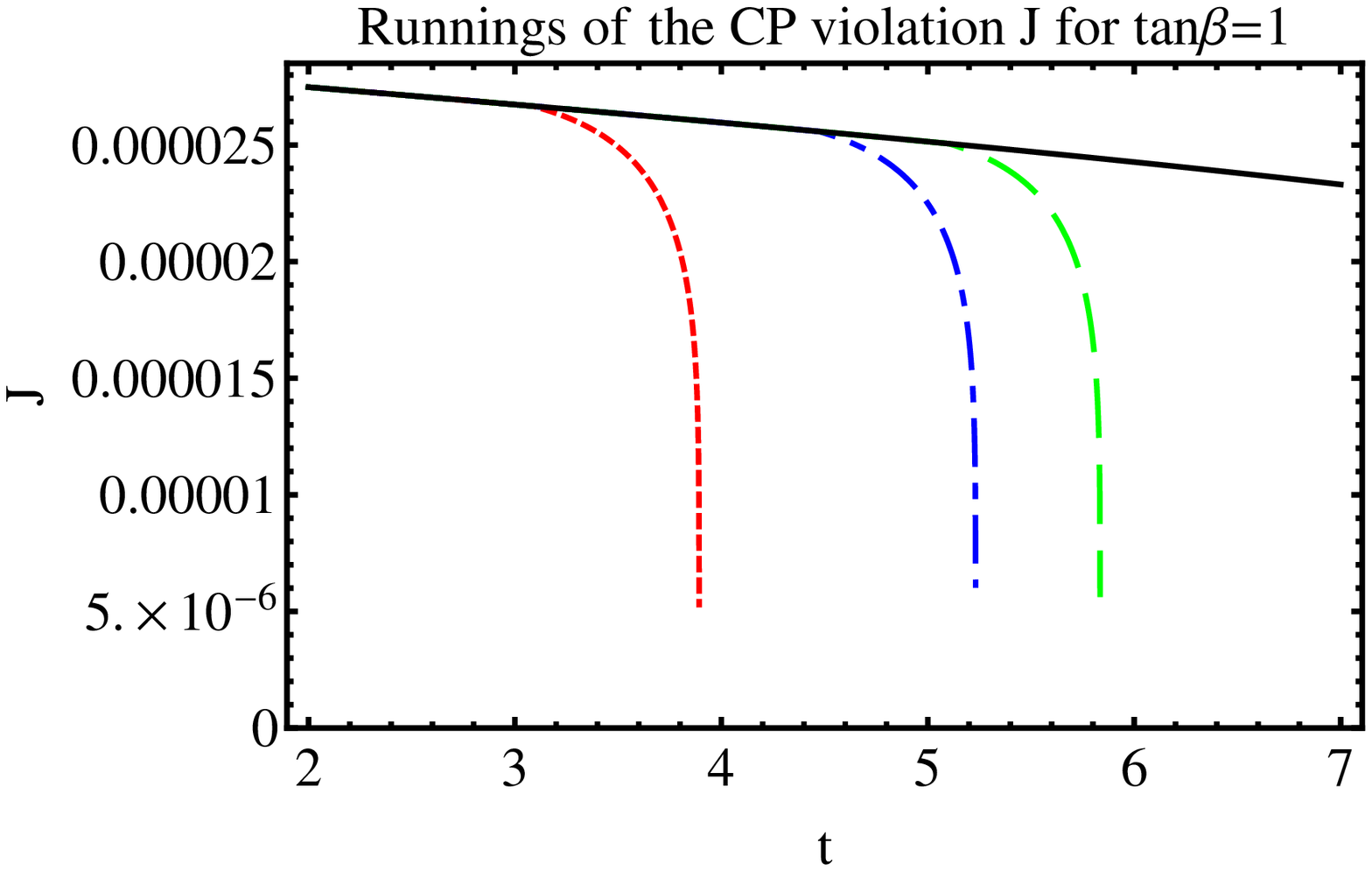}
\caption{\sl (Colour online) The CMK matrix elements $|V_{us}|$ (top left), $|V_{ub}|$ (top right), $|V_{cb}|$ (bottom left) and the Jarlskog parameter $J$ (bottom right) as functions of the scale parameter $t$ for $\tan\beta=1$. All matter fields are in the bulk for a variety of compactification scales: $R^{-1}$ = 2 TeV (red, dotted line), 8 TeV (blue, dot-dashed line), and 15 TeV (green, dashed line).}
\label{fig:5}
\end{center}
\end{figure}

\begin{figure}[th]
\begin{center}
\includegraphics[width=0.45\textwidth]{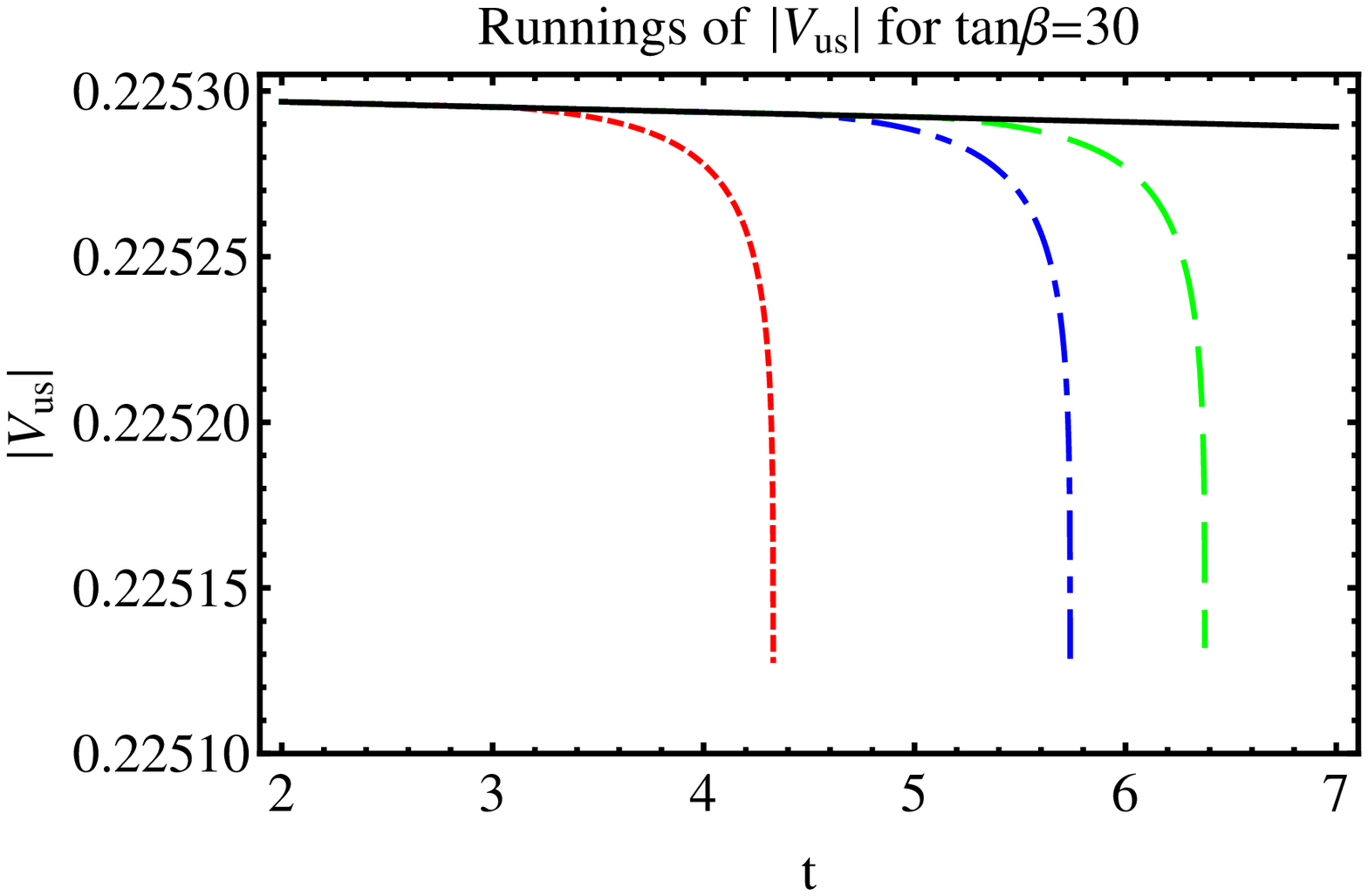}\hspace{0.05\textwidth}
\includegraphics[width=0.45\textwidth]{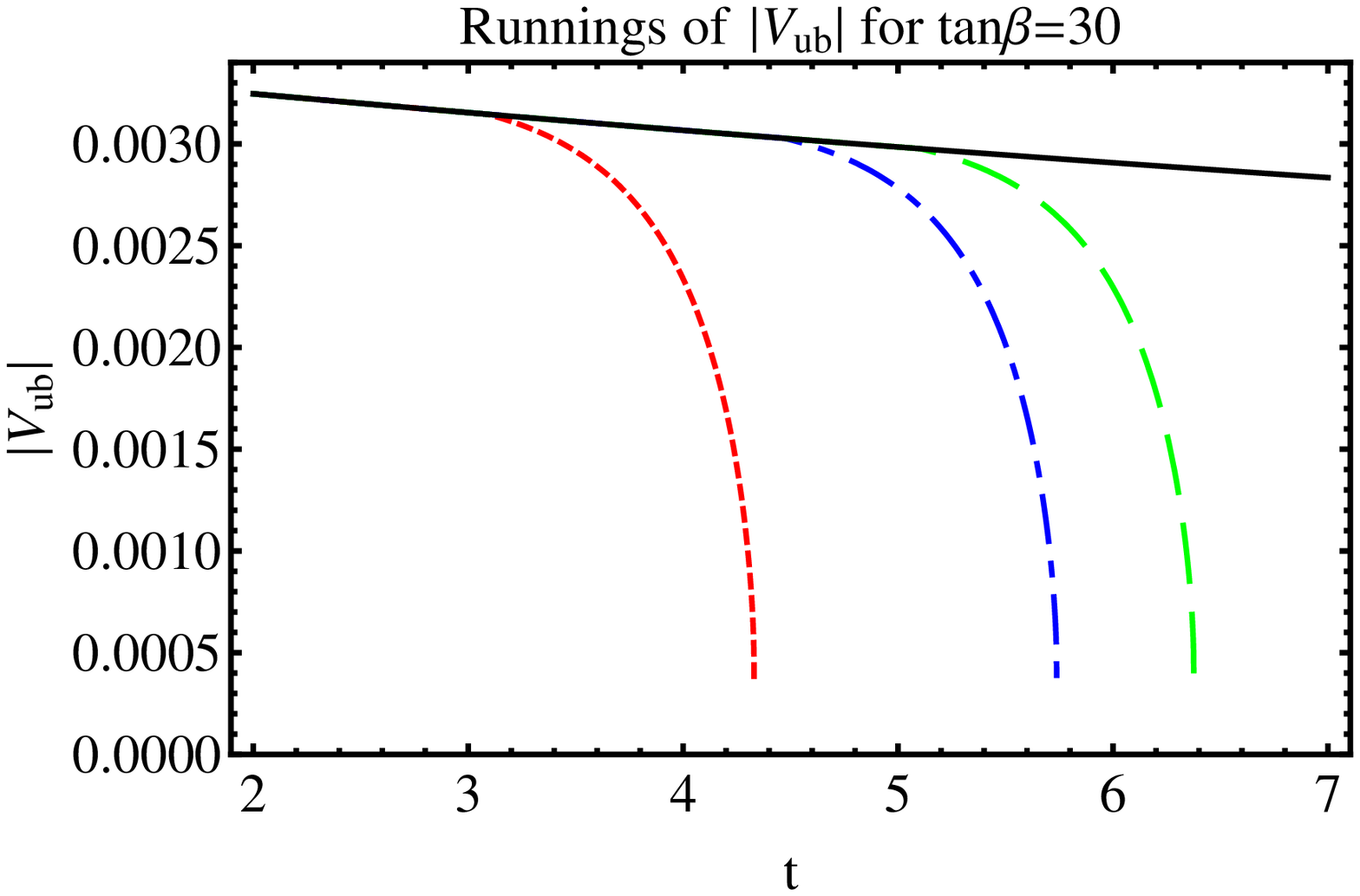}
\includegraphics[width=0.45\textwidth]{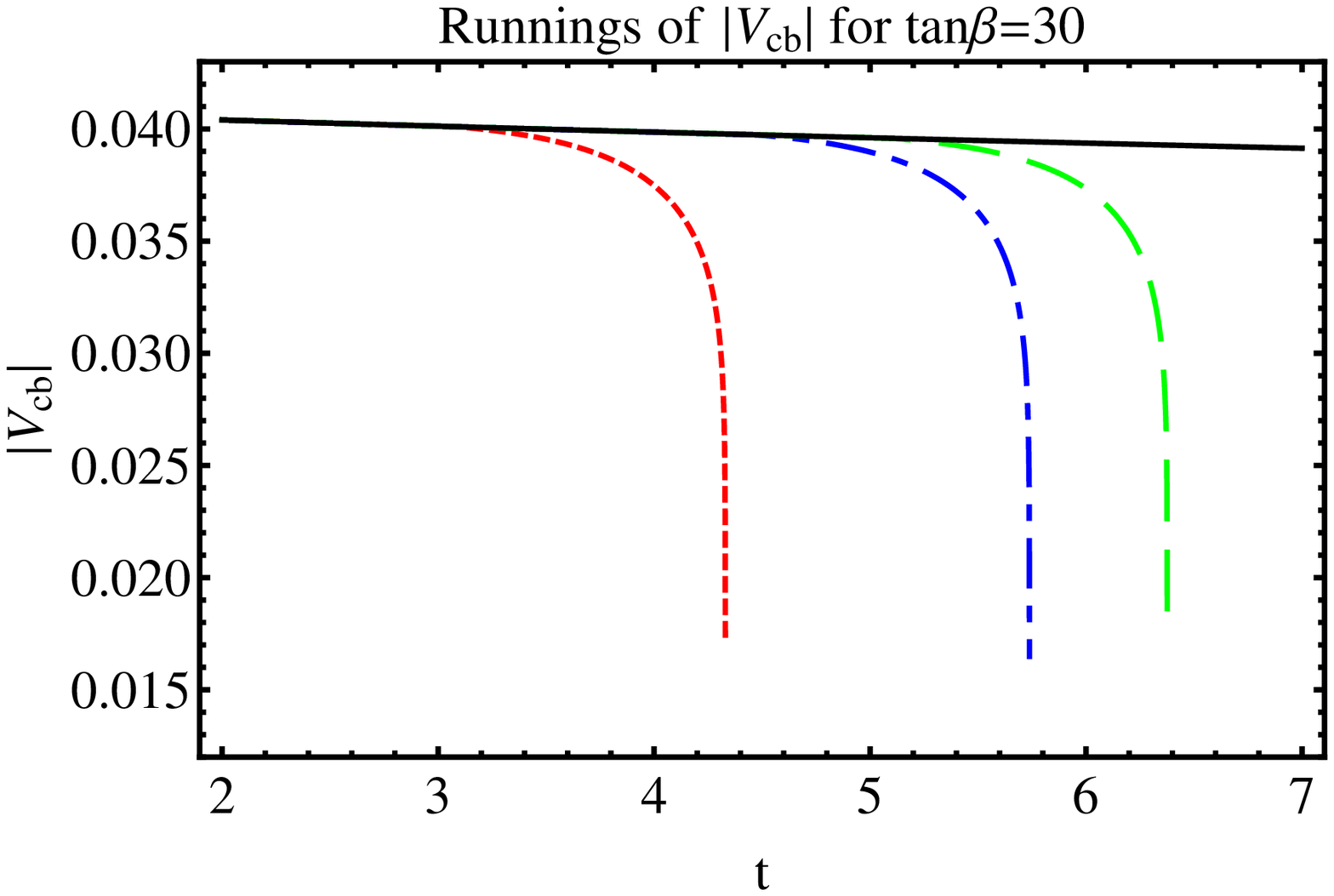}\hspace{0.05\textwidth}
\includegraphics[width=0.45\textwidth]{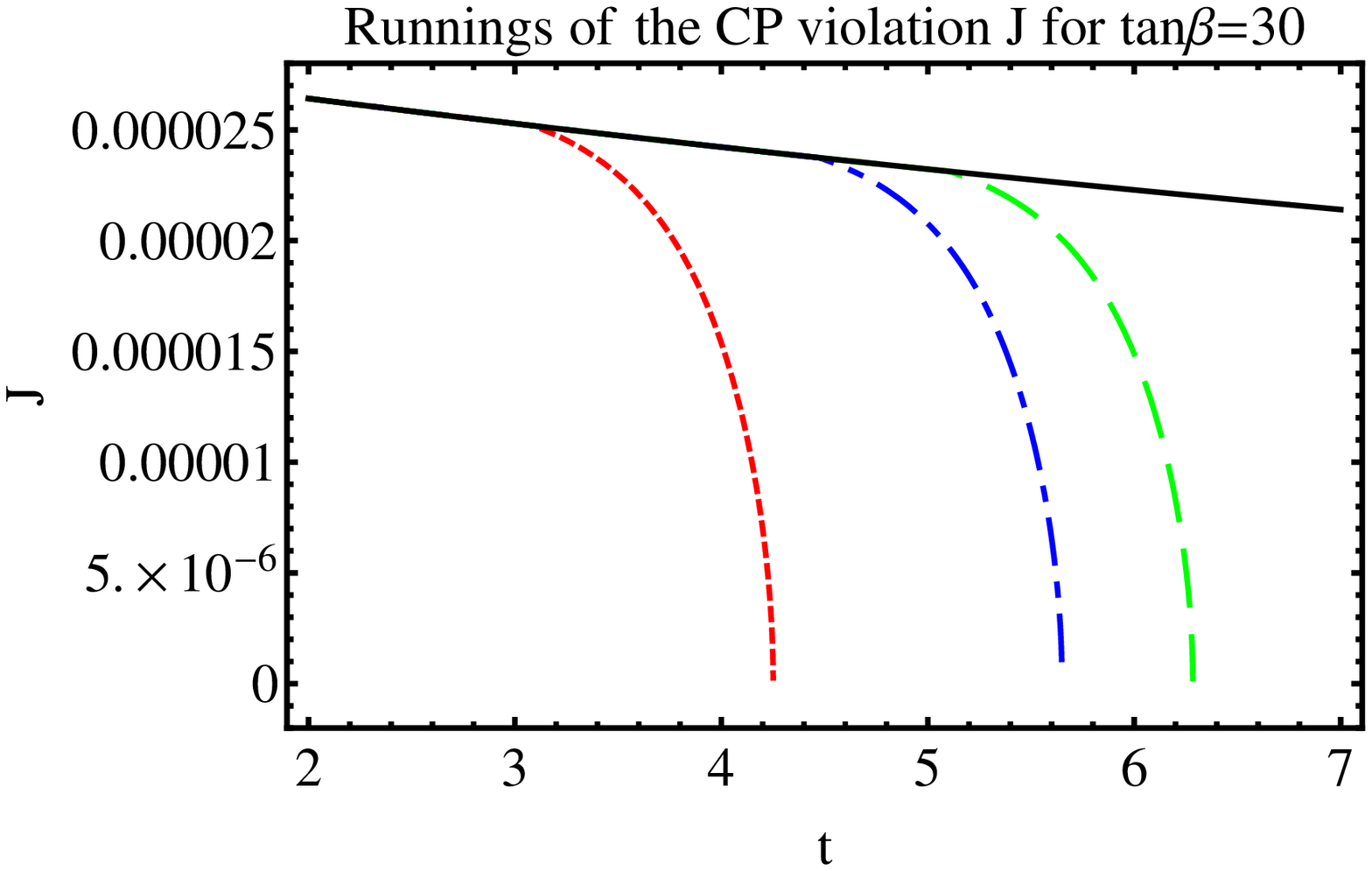}
\caption{\sl (Colour online) The CMK matrix elements $|V_{us}|$ (top left), $|V_{ub}|$ (top right), $|V_{cb}|$ (bottom left) and the
Jarlskog parameter $J$ (bottom right) as functions of the scale parameter $t$ for $\tan\beta=30$. All matter fields are in the bulk for a
variety of compactification scales: $R^{-1}$ = 2 TeV (red, dotted line), 8 TeV (blue, dot-dashed line), and 15 TeV (green, dashed line).}
\label{fig:6}
\end{center}
\end{figure}

\begin{figure}[t]
\begin{center}
\includegraphics[width=0.45\textwidth]{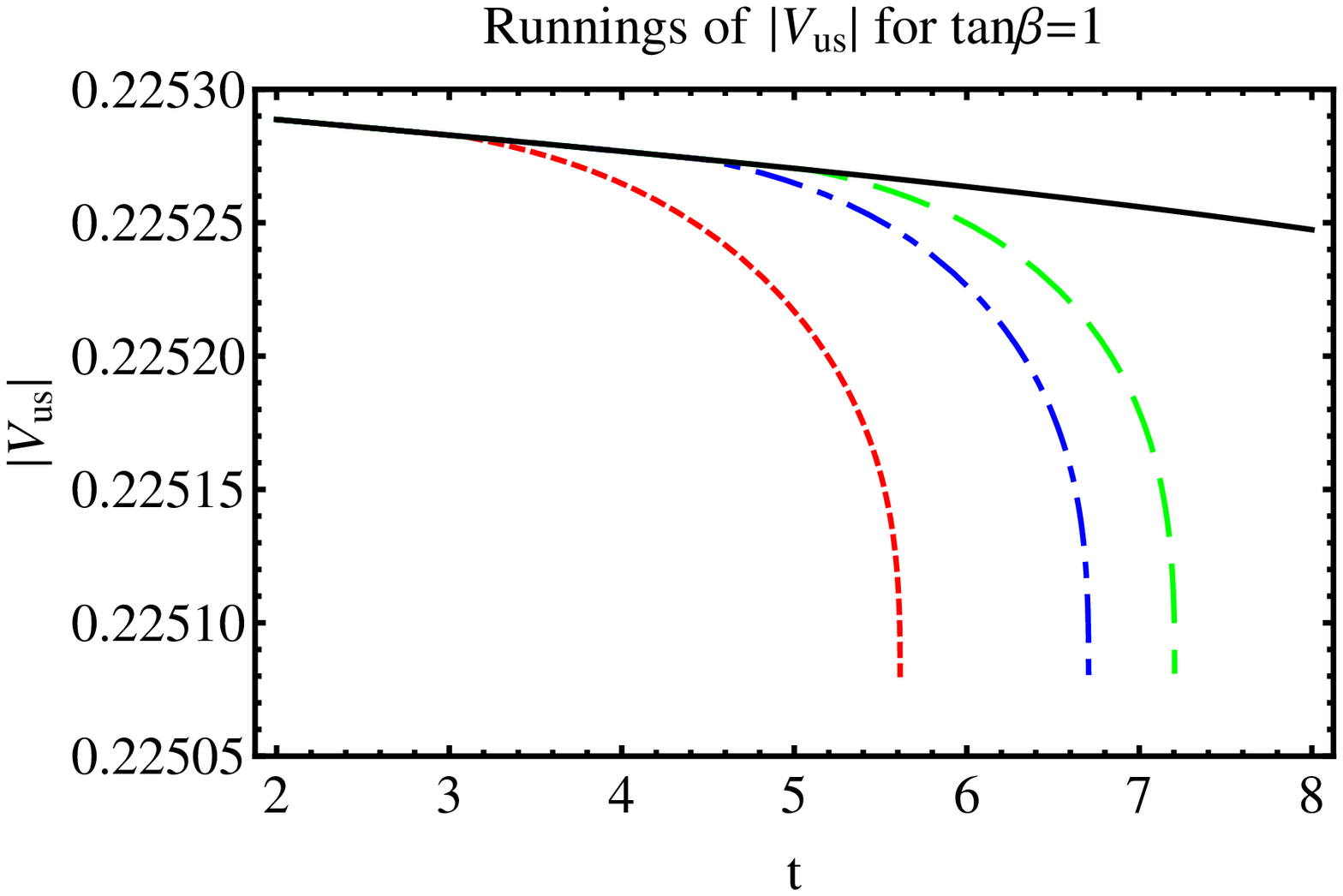}\hspace{0.05\textwidth}
\includegraphics[width=0.45\textwidth]{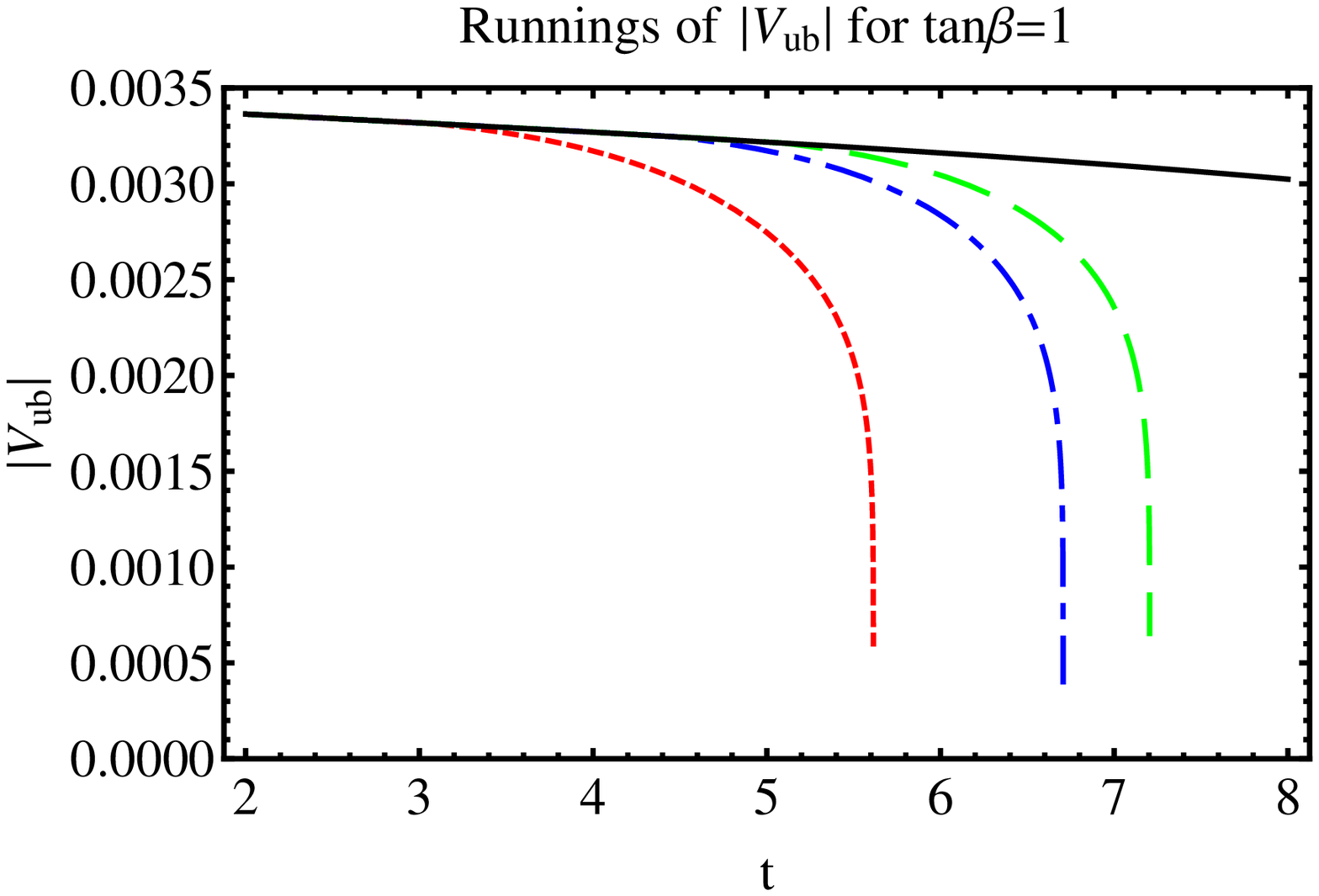}
\includegraphics[width=0.45\textwidth]{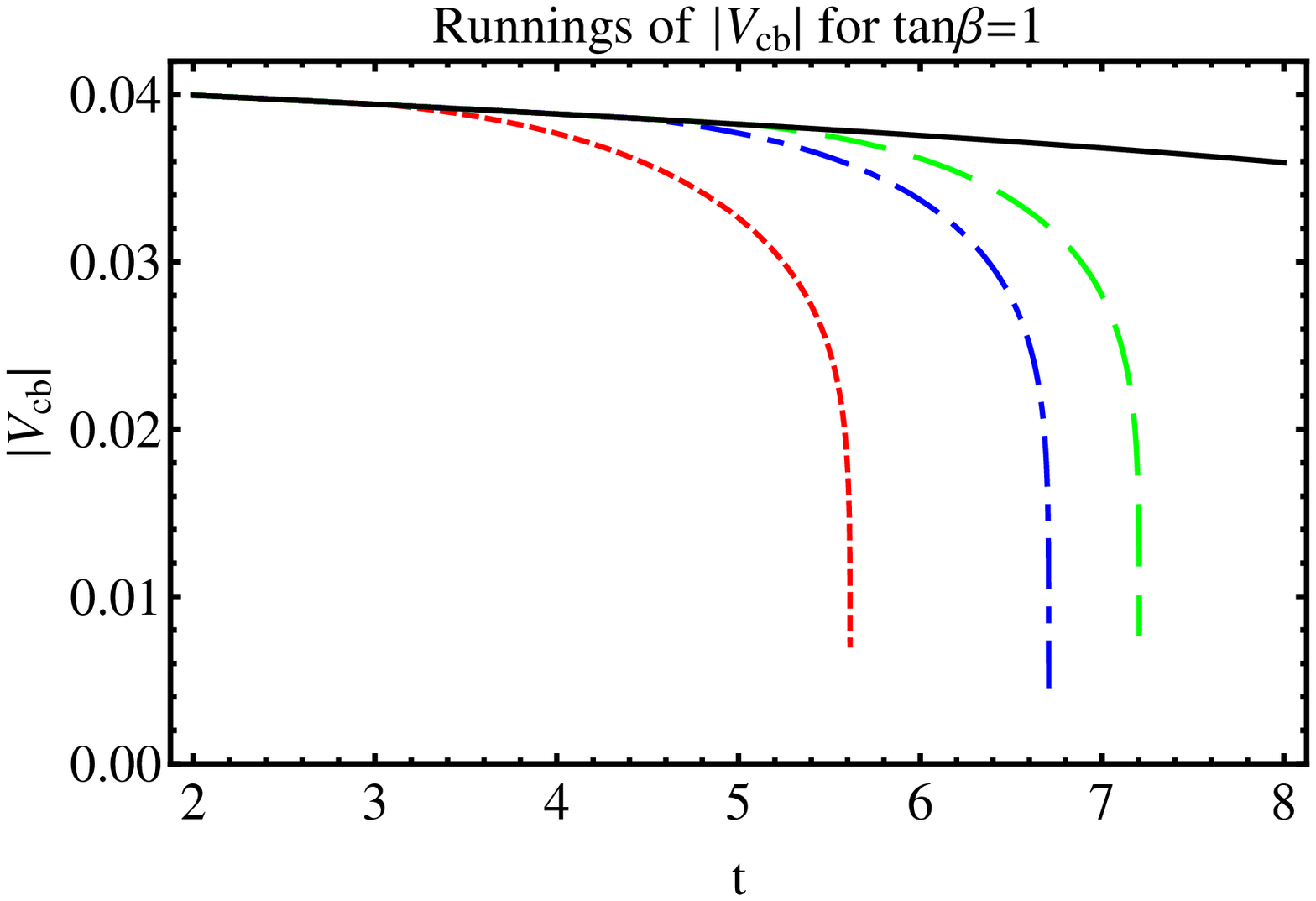}\hspace{0.05\textwidth}
\includegraphics[width=0.45\textwidth]{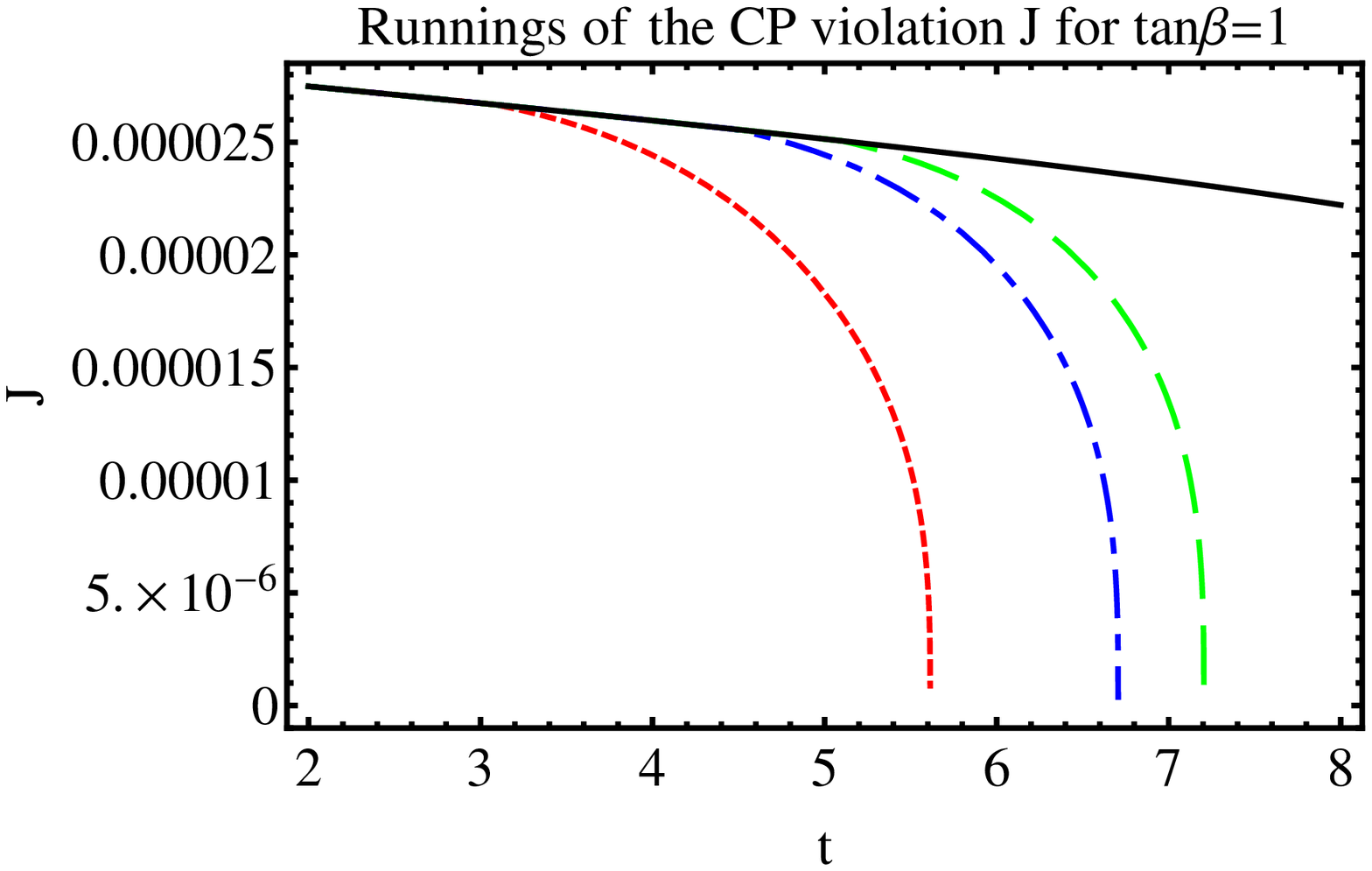}
\caption{\sl (Colour online) The CMK matrix elements $|V_{us}|$ (top left), $|V_{ub}|$ (top right), $|V_{cb}|$ (bottom left) and the
Jarlskog parameter $J$ (bottom right) as functions of the scale parameter $t$ for $\tan\beta=1$. All matter fields are on the brane for a
variety of compactification scales: $R^{-1}$ = 2 TeV (red, dotted line), 8 TeV (blue, dot-dashed line), and 15 TeV (green, dashed line).}
\label{fig:7}
\end{center}
\end{figure}

\begin{figure}[th]
\begin{center}
\includegraphics[width=0.45\textwidth]{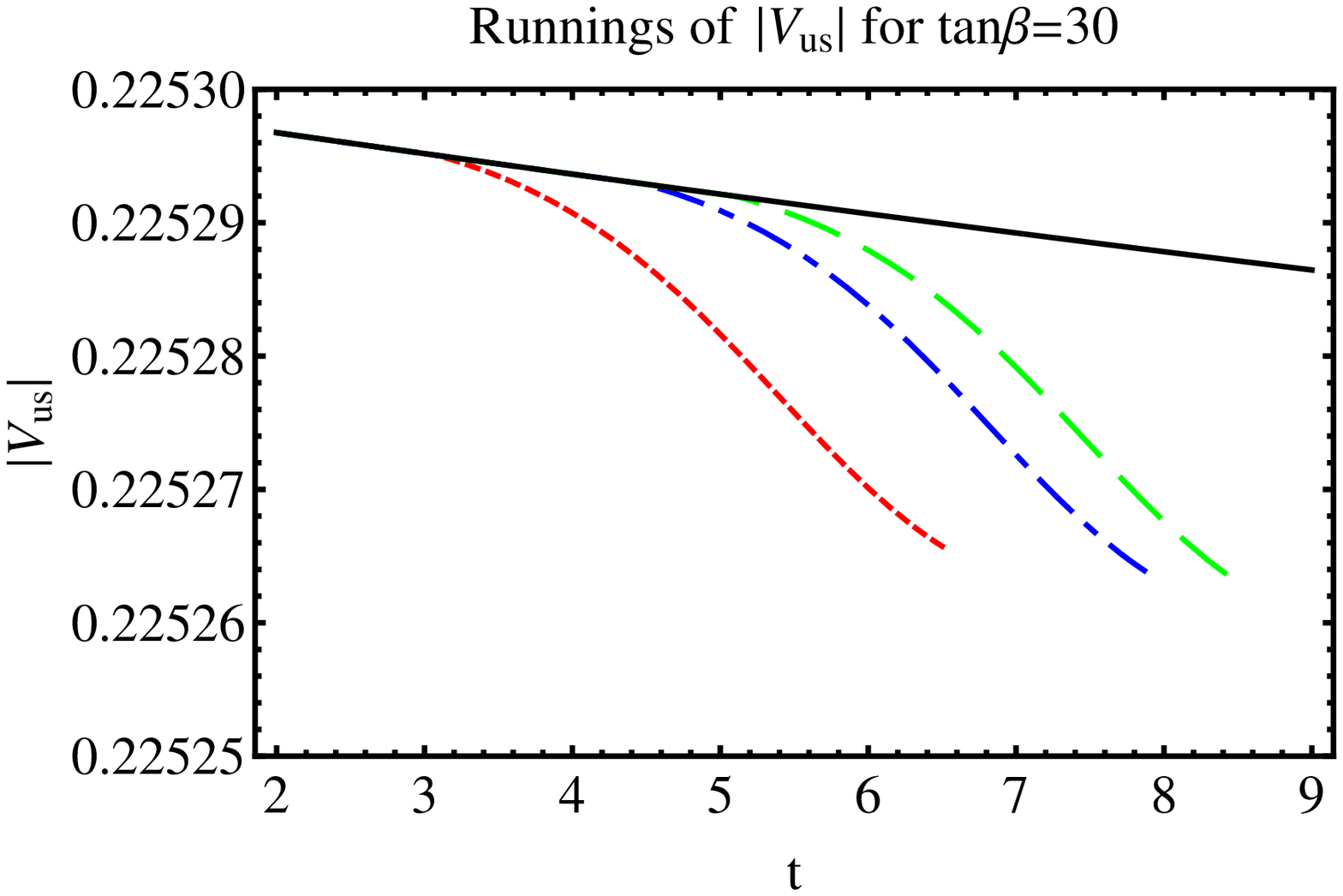}\hspace{0.05\textwidth}
\includegraphics[width=0.45\textwidth]{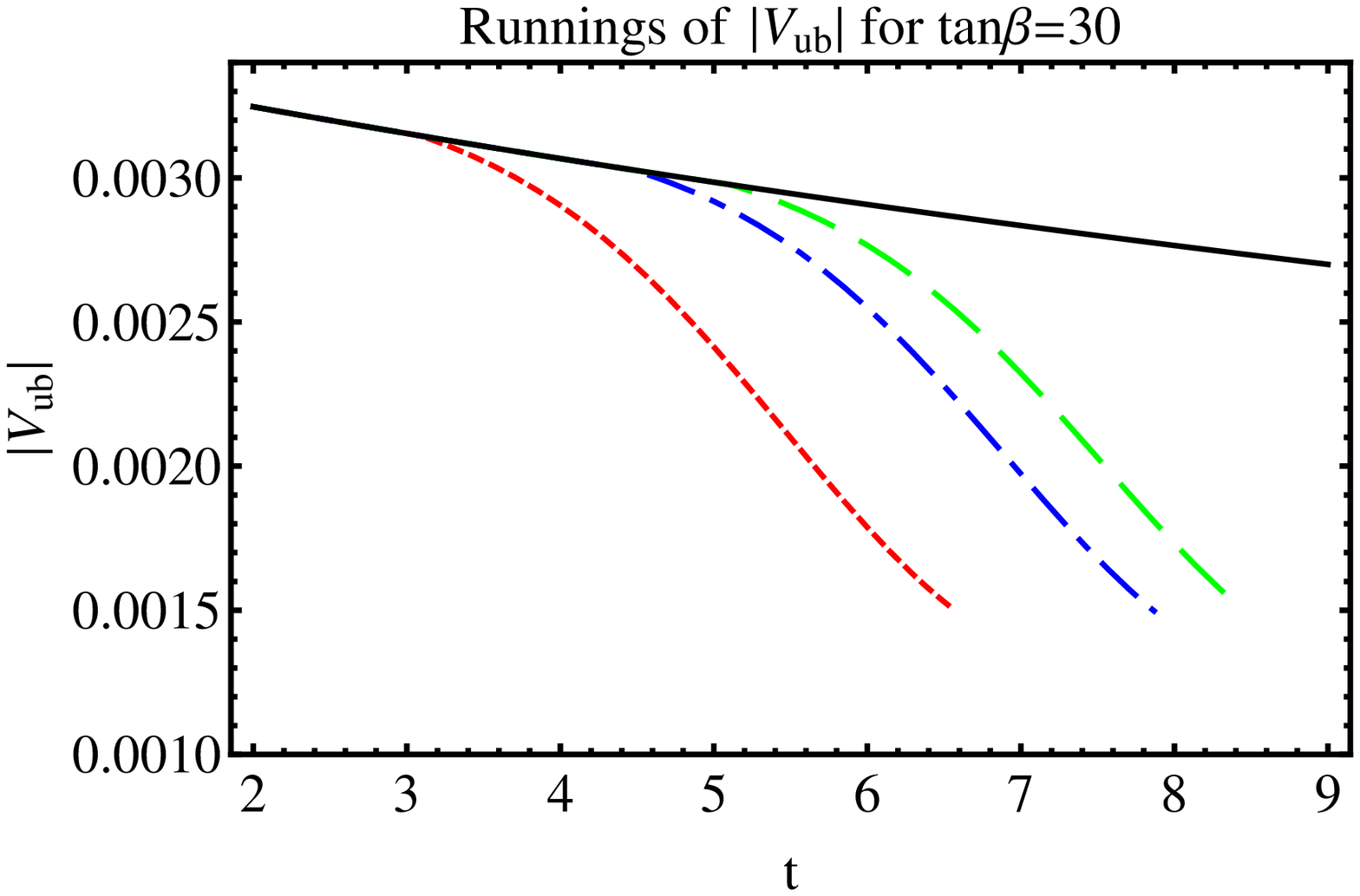}
\includegraphics[width=0.45\textwidth]{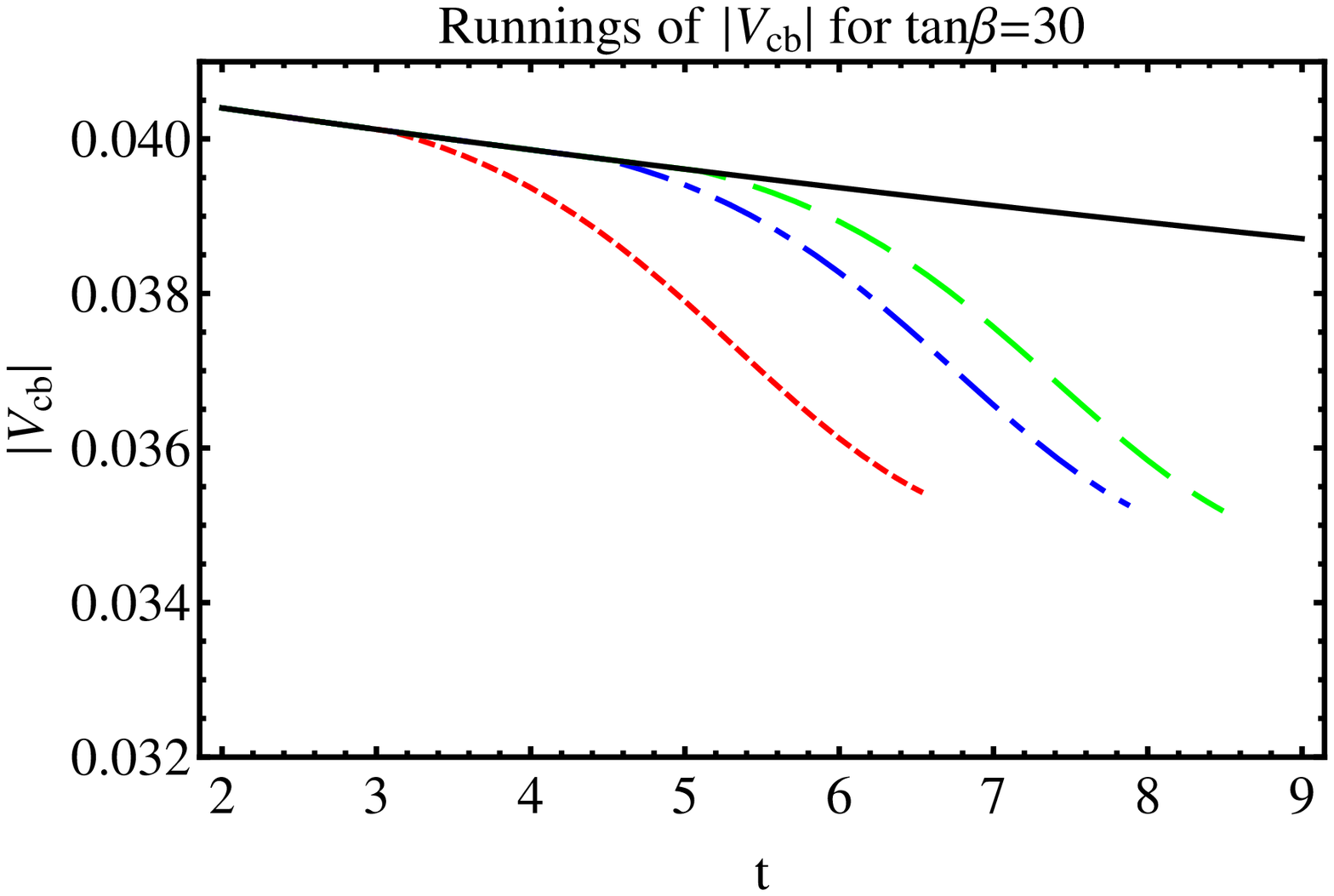}\hspace{0.05\textwidth}
\includegraphics[width=0.45\textwidth]{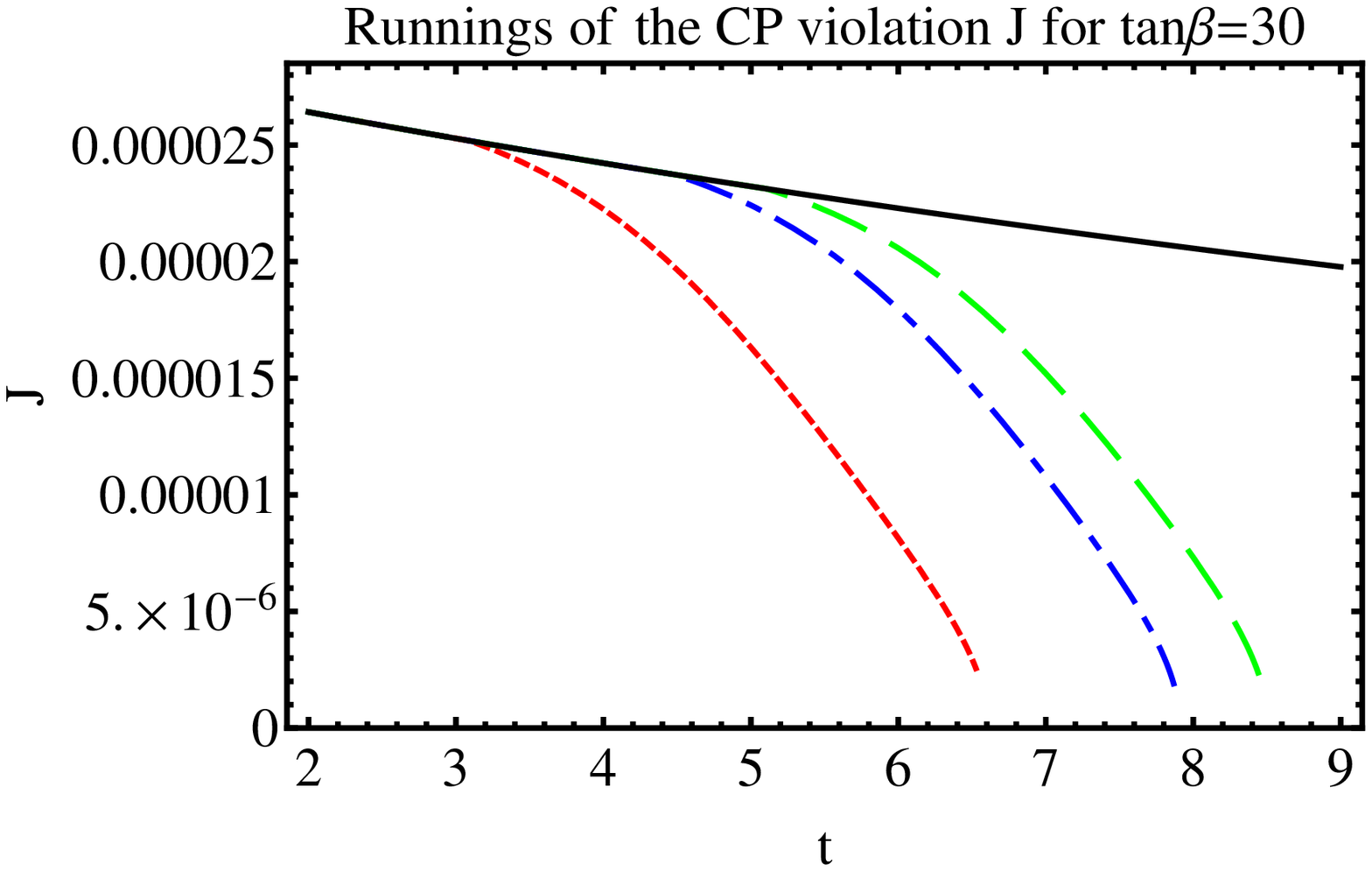}
\caption{\sl (Colour online) The CMK matrix elements $|V_{us}|$ (top left), $|V_{ub}|$ (top right), $|V_{cb}|$ (bottom left) and the
Jarlskog parameter $J$ (bottom right) as functions of the scale parameter $t$ for $\tan\beta=30$. All matter fields are on the brane for a
variety of compactification scales: $R^{-1}$ = 2 TeV (red, dotted line), 8 TeV (blue, dot-dashed line), and 15 TeV (green, dashed line).}
\label{fig:8}
\end{center}
\end{figure}

 We next turn our attention to the quark flavor mixing matrix, especially the complex phase of the CKM matrix which characterises CP-violating phenomena. This phenomena has been unambiguously verified in a number of $K - \bar{K}$ and $B - \bar{B}$ systems. Because of the arbitrariness in choice of phases of the quark fields, the phases of individual matrix elements of the $V_{CKM}$ are not themselves directly observable. Among these we therefore use the absolute values of the matrix element $|V_{ij}|$ as the independent set of rephasing invariant variables. Of the nine elements of the CKM matrix, only four of them are independent, which is consistent with the four independent variables of the standard parametrisation of the CKM matrix. For definiteness we choose the $|V_{ub}|$, $|V_{cb}|$, $|V_{us}|$ and the Jarlskog rephasing invariant parameter $J = \mathrm{Im} V_{ud}V_{cs}V^*_{us} V^*_{cd}$ as the four independent parameters of $V_{CKM}$.  In Figs.\ref{fig:5}-\ref{fig:6} we plot the energy dependence of these four variables from the weak scale all the way up to the high energy scales for different values of compactification radii $R^{-1}$ for the universal 5D MSSM case, and in Figs.\ref{fig:7}-\ref{fig:8} for the brane localized matter fields case. In these sets of pictures we consider two indicative choices of $\tan\beta$, that of $\tan\beta = 1$ and $\tan\beta = 30$.


\par The running of the CKM matrix is governed by the terms related to the Yukawa couplings. These Yukawa couplings are usually very small, except for the top Yukawa coupling (which could give a sizeable contribution). The CKM matrix elements $V_{us} \simeq \theta_{12}$, $V_{ub} \simeq \theta_{13} e^{-i \delta}$, $V_{cb} \simeq \theta_{23}$, can be used to observe the Cabibbo angle, $ \theta_{13}$ and $\theta_{23}$. For the universal 5D MSSM scenario we plot their evolution in Figs.\ref{fig:5}-\ref{fig:6}, and find that they decrease with the energy scale in a similar manner regardless of whether $\tan\beta$ is small or large. However, for a large initial value of $f_t$ (small $\tan\beta$), the mixing angles have a more rapid evolution and end in the regime where the top Yukawa diverges and develops a singularity. Similarly, the Jarlskog parameter also decreases quite rapidly once the initial KK threshold is passed. However, when $\tan\beta$ is large, we have a relatively longer distance between the initial and terminating energy track, the evolution of $J$ can be driven towards zero or even further. Besides, as can be seen explicitly in Ref. \cite{Babu:1987im}, the beta functions of the evolution equations of the CKM elements are up to the third order in the CKM elements, which are comparably smaller than that of Jarlskog parameter's quadratic dependence on the CKM elements. This fact then leads to the relatively large variation of J with the increase of energy. Quantitatively we observe from these plots the following; the decrease in the values of $V_{us}$ is not sizeable, whilst for $V_{ub}$ and $V_{cb}$ their variations change by more than 50\%. Furthermore, for $\tan\beta=30$, the Jarlskog parameter drops almost to zero, which sets the effect of the SM CP violation to being very small. Note, however, that in a supersymmetric theory other sources of CP violation beyond the SM ones are typically present, therefore only a complete and detailed study of a specific model would allow us to establish the strength of the CP violating effects.

\par For the case of the matter fields being constrained to the brane, in Figs.\ref{fig:7}-\ref{fig:8} we observe that the evolutions of these mixing angles and CP violation parameter are decreasing irrespective of whether the top Yukawa coupling grows or not. For small $\tan\beta$ we see similar evolution behaviours for these parameters as in the bulk case. The decreases in the values of $V_{us}$, $V_{ub}$, $V_{cb}$ and $J$ are much steeper, due to rapid growth of the top Yukawa coupling near the singular point. However, as $\tan\beta$ becomes larger, e.g. $\tan\beta=30$, the top Yukawa coupling evolves downward instead of upward. The decreases in these CKM parameters then becomes much milder towards the unification scale; though the reduction to effectively zero in the Jarlskog parameter persists. As a result, for the brane localised matter field scenario, it is more desirable to have a large $\tan\beta$ for theories that are valid up to the gauge coupling unification scale.


\section{Conclusions}\label{sec:5}

\par In summary, for the two 5D MSSM scenarios with matter fields in the bulk or on the brane, we have performed the numerical analysis of the evolution of the various parameters of the CKM matrix, and both cases give us a scenario with small or no quark flavor mixings  at high energies, especially for the mixings with the heavy generation. The evolution equations which relate various observables at  different energies, and also allow the study of their asymptotic behaviours, are particularly important in view of testing the evolution  of the Yukawa couplings. In the universal 5D MSSM model, the evolution of these CKM parameters have a rapid variation prior to reaching a cut-off scale where the top Yukawa coupling develops a singularity point and the model breaks  down. For the brane localised matter fields model, we can only observe similar behaviours for small values of $\tan\beta$, while for  large $\tan\beta$, the initial top Yukawa coupling becomes smaller, the gauge couplings then play a dominant role during the evolution  of the Yukawa couplings, which cause the Yukawa couplings to decrease instead of increasing. As such the variations of these CKM  parameters have a relatively milder behaviour, and the theory is valid up the gauge coupling unification scale.


\end{document}